\documentclass[manuscript]{aastex62}
\usepackage{makeidx}

\usepackage[inline]{enumitem}

\submitjournal{Icarus}

\shorttitle{Asteroids were born bigger}
\shortauthors{Ribeiro et al.}

\begin{document}

\title[Asteroids were born bigger]{Asteroids were born bigger: An implication of surface mass ablation during gas-assisted implantation into the asteroid belt }

\correspondingauthor{Rafael Ribeiro de Sousa}
\email{r.sousa@unesp.br}

\author[0000-0002-0786-7307]{Rafael Ribeiro de Sousa}
\affiliation{S\~ao Paulo State University, UNESP,Campus of Guaratinguet\'a, Av. Dr. Ariberto Pereira da Cunha, 333 - Pedregulho, Guaratinguet\'a - SP, 12516-410, Brazil}

\author{Andre Izidoro}
\affil{Department of Earth, Environmental and Planetary Sciences, 6100 Main MS 126,  Rice University, Houston, TX 77005, USA}

\author{Rogerio Deienno}
\affil{Southwest Research Institute, 1050 Walnut St. Suite 300, Boulder, CO 80302, USA}

\author{Rajdeep Dasgupta}
\affil{Department of Earth, Environmental and Planetary Sciences, 6100 Main MS 126,  Rice University, Houston, TX 77005, USA}

\begin{abstract}
\small

The origins of carbonaceous asteroids in the asteroid belt is not fully understood. The leading hypothesis is that they were not born at their current location but instead implanted into the asteroid belt early in the Solar System history. In this study, we investigate how the migration and growth of Jupiter and Saturn in their natal disk impact nearby planetesimals and subsequent planetesimal implantation into the asteroid belt. Unlike traditional studies, we account for the effects of surface ablation of planetesimals caused by thermal and frictional heating between the gas-disk medium and planetesimal surface, when planetesimals travel through the gas disk. We have performed simulations considering planetesimals of different compositions as water-ice rich planetesimals (composed of more than 80\% water by mass), water-ice poor planetesimals (relatively dry and enstatite-like), organic-rich planetesimals, and fayalite-rich planetesimals. Our findings indicate that, regardless of the migration history of the giant planets, water-ice rich, organic-rich, and fayalite-rich planetesimals implanted into the asteroid belt generally experience surface ablation during implantation in the asteroid belt, shrinking in size. On the other hand, planetesimals with enstatite-like compositions were inconsequential to surface ablation, preserving their original sizes. By assuming an initial planetesimal size-frequency distribution (SFD), our results show that -- under the effects of surface ablation -- the planetesimal population implanted into the asteroid belt shows a SFD slope slightly steeper than that of the initial one. This holds true for all migration histories of the giant planets considered in this work, but for the Grand-Tack model where the SFD slope remains broadly unchanged. Altogether, our results suggest that the largest C-type asteroids in the asteroid belt may have been born bigger. High-degree surface ablation during implantation into the asteroid belt may have even exposed the cores of early differentiated C-type planetesimals.

\end{abstract}


\section{Introduction \label{Sec:Introduction}}

The main asteroid belt is primarily composed
of two taxonomic classes of asteroids, i.e. silicate-rich asteroids (S-types) and carbon-rich asteroids (C-types). The orbital distributions of S- and C-type asteroids broadly overlaps, yet S-types mainly populate
the inner regions ($2.0 < a < \sim 2.5$ au), while C-types more commonly populate the central and outer regions ($2.5 < a < \sim 3.4$ au;
\cite{1982Sci...216.1405G,1985LPI....16..432K,2013Icar..226..723D,2018SSRv..214...36A}). S- and C-type
asteroids are correlated spectroscopically with different
types of meteorites \citep{burbine2002asteroids}. S-types are associated with non-carbonaceous (NC) chondrite meteorites, while C-types are associated with carbonaceous (CC) chondrite meteorites.

It has now been extensively documented that a distinct stable-isotopic composition -- for several volatile and non-volatile chemical elements as O, Ti, Ni, Mo, Nd, Cr, V, N, etc  -- exists between the NC and CC classes of meteorites~\citep{WARREN201193,Kruijer2017,2020NatAsK,2021NatAsG,2022ApJG}. The isotopic compositions of NC and CC meteorites suggest they may have accreted from distinct dust and pebble reservoirs that remained disconnected for 2-4 million years~\citep{Kruijer2017}. This supports the hypothesis that the early solar system was  composed of at least two  chemical reservoirs that were physically isolated from each other either due to the formation of Jupiter \citep{2020NatAsK,hoppetal22}, or due to the existence of pressure bumps in the disk~\citep{2020NatAsB,2021NatAs.tmp..262I}. NC parent bodies are believed to have formed in the inner parts of the solar system, potentially inside the water snowline\footnote{The location in the gaseous disk where water condenses as ice.} as suggested by their low water content. CC parent bodies are believed to have formed farther out  -- most likely beyond  the water snowline \citep{Kruijer2017,raymond17,2021NatAs.tmp..262I}, as suggested by their relatively higher water contents~\citep[e.g.][]{2023arXiv230202674I}.

Different mechanisms and models have been proposed to explain the  broadly overlapping orbital distributions of  C- and S-type asteroids in the asteroid belt. We dedicate this work to revisit models for the origins of C-type asteroids in the asteroid belt. For studies focusing on the origin of S-type asteroids, we refer the reader to \cite{2005Icar..179...63B,Walsh2011,raymond17,2021NatAs.tmp..262I}.

The leading hypothesis for the origin of C-type asteroids is that they were implanted into the asteroid belt early in the solar system history, but the very processes and mechanisms behind the implantation process remain unconstrained at the present time. In the so-called Grand-Tack model~\citep{Walsh2011}, Jupiter and Saturn are envisioned to have experienced inward-then-outward migration in the Sun's natal disk~\citep{MassetSnellgrove2001,pierens14}. The role of Jupiter and Saturn in the Grand-Tack model is to sculpt the formation of the terrestrial planets and the structure of the asteroid belt. Jupiter is assumed to have formed at about 3.5~au away from the Sun and  migrated down to 1.5~au, depleting the asteroid belt region and creating a narrow ring of planetesimals located at around 1 au. The subsequent accretion of terrestrial planets from a narrow ring of planetesimals is consistent with the terrestrial planets masses, in particular with the low-mass of Mars \citep{2009ApJ...703.1131H,Walshetal2011,izidoroetal14}. Saturn plays also an important role in this model. As Saturn grows and migrates inwards, it  eventually captures Jupiter in a mean motion resonance, which reverses their migration direction, bringing both giant planets from $\sim$1.5-2~au~\citep{Walshetal2011,brasseretal16} to close to their current locations. During the outward migration phase, Jupiter and Saturn cross the asteroid belt region once again.  The implantation of C-type asteroids into the asteroid belt in this scenario  takes place mostly during the outward migration phase of Jupiter and Saturn when these planets scatter planetesimals inwards ``filling up'' the initially depleted asteroid belt.

The Grand-Tack model invokes a very specific migration history for the giant planets in order to create a narrow ring of material, broadly reproducing the terrestrial planets, and implanting C-type asteroids in the asteroid belt. Yet, a Grand-Tack like evolution is not required to explain the inner solar system if planetesimals in the terrestrial region formed from a narrow ring of dust grains \citep{2009ApJ...703.1131H,izidoroetal14,2021NatAs.tmp..262I,morbidellietal2022}. The origin of C-type asteroids in the asteroid belt may be instead a simple consequence of Jupiter and Saturn's growth and/or {\it inward} migration in the Sun's natal disk~\citep{2017SciARaymondIzidoro}. When Jupiter's and Saturn's cores started runaway gas accretion~\citep{Pollacketal1996} they dynamically scattered any nearby planetesimals. Dynamical scattering takes place because the stability regions around growing planets changes as their masses increase~\citep{raymond17}. The mass of a growing giant planet increases by at least one order of magnitude in a few million of years (from  $\sim10M_{\oplus}$ cores  to $\sim 100M_{\oplus}$ planets).  Due to the effects of gas-drag \citep{Adachi1976}, a fraction of the planetesimals scattered by Jupiter and Saturn eventually reached the inner regions of solar system and were implanted into the asteroid belt~\citep{raymond17}.  This is a generic mechanism that is expected to operate whenever a gas giant rapidly grows in the presence of nearby planetesimals. In addition,  planet migration occurring during the process of giant planet growth, which is typically inward \citep{Kley2009}, also leads to planetesimal shepherding ~\citep{raymondetal06} and implantation of asteroids in the belt~\citep{raymond17,Deienno_2022}.

Although these different scenarios can broadly reproduce the inner Solar System architecture~\citep{izidororaymond18}, in particular the asteroid belt structure, a common drawback of these studies is that they, for simplicity, have neglected the effects of surface ablation on planetesimals during their implantation phase into the asteroid belt~\citep{raymond17,Deienno_2022}. When a planetesimal is scattered (or shepherded) on an eccentric orbit its surface can be heated due to friction with the gas medium. This is a consequence of collisions between the gas molecules (or atoms) and the planetesimal transferring some fraction of the collisional kinetic energy as heat~\citep{2015ApJ...806..203D}. A planetesimal can also be heated by absorbing radiation emitted by the local gas medium and, conversely, cooled down by the radiation emission from its surface. Depending on the planetesimal composition, size, and local disk temperature, material at the surface of the planetesimal can experience phase transition due to heating and  eventually be liberated into space in the form of gas. This process is known as \textit{surface ablation}. It has been particularly studied in the context of meteorite ablation during atmospheric entry on Earth and planetesimal ablation on the envelope of growing giant planets \citep{1924MWRv...52..488W, 1971NASSP.252..219B,2004A&A...418..751C,2015ApJ...806..203D,refId0,2021A&A...648A.112E}. Here we are interested in understanding how surface ablation would potentially change the size of planetesimals implanted into the asteroid belt during the Sun's natal disk phase. In this work, we neglect the effects of planetesimal-planetesimals collisions~\citep[see, for instance,][]{carterstewart20,Deienno_2022}.

The effects of surface ablation was investigated in the context of water-ice rich planetesimals growing at the edge of gaps carved by giant planets~\citep{2021A&A...648A.112E}. The authors found that water-ice rich planetesimals are easily destroyed due to the effects of surface ablation, if they are injected in regions within $10~{\rm au}$, provided the gaseous disk is still around. In this work, we are mainly interested in understanding how the initial size-frequency distribution (SFD) of planetesimals is affected when surface ablation is accounted for during the implantation of asteroids into the asteroid belt. To revisit this problem, we have performed high-resolution N-body numerical simulations modelling the implantation of planetesimals from the giant planet region into the asteroid belt. We have performed simulations considering different radial temperature  profiles, initial gas densities for the disk,  planetesimals sizes and compositions, and also  different migration histories of the giant planets. Our results suggest that regardless of the growth and migration histories of the giant planets, outer solar system C-type planetesimals (mostly water-ice and organic rich) are likely to have experienced significant surface ablation during gas-assisted implantation in the asteroid belt, unless their surface material is as resistant to thermal perturbations and frictional heating as in the case for enstatite-like minerals. Planetesimals that experienced surface ablation during the asteroid belt's implantation process tend to show a SFD slope slightly steeper than that of the initial population. This holds true regardless of the migration histories of Jupiter and Saturn, but in the Grand-Tack model where the SFD slope of implanted planetesimals remains unchanged. Our results suggest that the largest C-type objects in the asteroid belt -- those that avoided substantial collisional evolution after implantation (e.g., D$>$100 km) -- were probably born bigger and had their sizes (significantly) reduced  during the implantation process. High-level surface ablation  may have even exposed the cores of differentiated early formed planetesimals.

This paper is structured as follows. In section \ref{Sec:Methods} we describe our methods. In Section \ref{Sec:Numerical_Simulations}, we present our numerical simulations, and in Section \ref{Sec:Results} we present our results. Finally, in Section \ref{conclusion}, we summarize our main findings and discuss our results.

\section{Methods}\label{Sec:Methods}

\subsection{Giant Planet Migration Histories}
\label{nominalpattern}
We have performed simulations modelling the implantation of planetesimals in the asteroid belt testing three different migration and growth histories for Jupiter and Saturn. These three scenarios are designed to capture a range of plausible evolutions of Jupiter and Saturn in a gaseous disk. We have performed simulations mimicking: i) the Grand-Tack evolution (thereafter referred to as ``GT''); ii) in-situ growth of Jupiter and Saturn at 5.4 and 7.25~au, respectively (thereafter referred to as ``insituJSrapid'') ; and iii) growth and rapid inward migration of Jupiter and Saturn (thereafter referred to as ``migJSrapid''). Scenario i) is inspired by the simulations of \citet{Walsh2011} and Scenarios ii) and ii) are inspired by simulations of \cite{raymond17}. We start by describing how we model the migration of Jupiter and Saturn as in the GT model.

To model the GT evolution of Jupiter and Saturn, we follow \citet{Walshetal2011}. The underlying 1D gaseous protoplanetary disk used as input in our simulations is the output of hydrodynamical simulations modelling the evolution of Jupiter and Saturn in a gaseous 2D disk~\citep{MorbidelliCrida2007}. We account for the gaps created by Jupiter and Saturn and we rescale the gas surface density profiles as the planets migrate inward and outward \citep{Walshetal2011}. The initial gas surface density of the disk in the hydrodynamical simulation is given by $\Sigma(r)=1700$~g/cm$^2$ at $1\sim$ au \citep{Weidenschilling1977,Hayashi1981,Walshetal2011}. We force the giant planets to migrate by applying artificial forces that mimic the effects of gas-disk tidal interactions~\citep{MassetSnellgrove2001,pierens14}. In our simulations, Jupiter is assumed completely formed and initially placed at 3.0 au. Jupiter is forced to  migrate inwards in a timescale of $10^{5}$ years, while Saturn's 30 $M_{\oplus}$ core remains fixed at 4.5 au and grow linearly with time. When Jupiter reaches 1.5 au, Saturn reaches 60 $M_{\oplus}$, undergoes a rapid inward migration phase,  and becomes captured in a 3:2 mean motion resonance (MMR) with Jupiter. At this point, the migration direction is reversed, and the two gas giants migrate outward together, with Jupiter reaching a final location of $\sim5.4$ au and Saturn at $\sim7.25$ au \citep{Nesvorny2012,Deiennoetal2017}.

Our simulations corresponding to the insituJSrapid and migJSrapid scenarios use the same gas disk profiles of \citet{raymond17}. The initial gas disk surface density profile is given by $\Sigma(r) = 4000(\frac{r}{au})^{-1} g/cm^2$. We mimic the growth of the giant planets by linearly increasing their masses over a timescale of $\tau_{grow} = 10^{5}$ years.  In the insituJSrapid scenario, Jupiter starts at $\sim 5.4$ au, with Saturn initially at $\sim 7.25$ au in a 3:2 MMR with Jupiter~\citep{raymond17}. In the migJSrapid scenario, the cores of Jupiter and Saturn are assumed to have formed farther out at $10$ and $15$ au, respectively, and migrated inwards to $\sim 5.4$ and $\sim 7.25$ au respectively, locked in a 3:2 MMR. In simulations following the insituJSrapid and migJSrapid scenarios,  Jupiter and Saturn initially start as 3 $M_{\oplus}$ cores. Following \citet{raymond17}, as the giant planet cores grow, we account for planet-induced gaps by  linearly interpolating between the initial power-law surface density profile and a disk with a fully open-gap sculpted by Jupiter (or Jupiter and Saturn).  Both planets start to grow simultaneously at the beginning of the simulation. All the simulations are conducted for 1 Myr, and we mimic the long term disk evolution of the disk, for the three sets of simulations, by considering that the gas surface density decays exponentially with an e-fold timescale of 100 Kyr.

\subsection{Disk temperature profile}
\label{temperature}

The formation of Jupiter and Saturn in the gaseous disk have the potential to excite spiral shock waves, which can impact the overall thermal structure of the disk. Both gap opening and shock heating can lead to the displacement of the snowline \citep{2020A&A...633A..29Z}. In our simulations, we test the effect of the disk temperature by considering two scenarios for the  disk midplane temperature. We have performed simulations with disk radial temperature profiles \citep{1997ApJ...490..368C} given by  $T(r) = 150 K (r/au)^{-3/7}$, which we refer to as \textit{cold disk scenario}, and $T(r) = 250 K (r/au)^{-3/7}$, which we refer to as \textit{hot disk scenario}. In our cold disk, the snowline is initially at 1~au, which is well inside the inner edge of the main asteroid belt. In our hotter disk, it is initially at 3.3~au, which is roughly at the outer edge of the main asteroid belt. This allow us to probe the effects of the disk temperature for surface ablation for two end-member disk scenarios.

\subsection{Gas drag effects}
\label{gas}

The aerodynamic gas drag force on a particle moving in a gas disk can be expressed as function of its shape, size, velocity, and gas conditions. In the particular case of spherical
object with radius $R$, the drag force is expressed as \citep{Grishin2015}:
\begin{equation}
\vec{F}_D= - \frac{1}{2} C_D \pi R^2 \rho_g v_{rel} \vec{v}_{rel} ,
\end{equation}
where $C_D$ is the drag coefficient, and $\vec{v}_{rel}$ is the relative velocity vector between the gas and the planetesimal.
The drag coefficient for a spherical object is a function of the Reynolds number ($R_e$), the Mach Number ($M$) and Knudsen number ($K$).
We calculate the drag coefficient following \citet{Brasseretal2007}, where they estimated
the values of $M$, $K$ and $R_e$ as a function of planetesimal's size and velocity. For planetesimals sufficiently massive, in  the mass range of $m\sim {10}^{21}-{10}^{25}$~g, gas dynamical friction (GDF) may also play an important role, potentially  dominating over gas-drag effects \citep{Grishin2015}. Our simulations also account for the effects of GDF. The GDF force can be calculated as \citep{Grishin2015}:
\begin{equation}
\vec{F}_{GDF}= - \frac{4 \pi G^2m^2 \rho_g}{v_{rel}^3} \vec{v}_{rel} I(M) ,
\end{equation}
where $I(M)$ is a dimensionless factor depending on the Mach number ($M$), and given by:
\begin{itemize}
 \item If $M<1$ then:
\begin{equation}
I(M)=\frac{1}{2}\ln{\left(\frac{1+M}{1-M}\right)} -M
\end{equation}
 \item If $M>1$ then:
\begin{equation}
 I(M)=\frac{1}{2}\ln{\left(1- \frac{1}{M^2}\right)+ \ln{\left(\frac{v_{rel}t}{R}\right)}}
\end{equation}
\end{itemize}

\subsection{Surface Mass ablation}
\label{surfaceablation}

We model the effects of surface ablation on planetesimals following previous studies \citep{1924MWRv...52..488W, 1971NASSP.252..219B,2004A&A...418..751C,2015ApJ...806..203D,refId0,2021A&A...648A.112E}. The  mass ablation rate of a planetesimal with chemical composition ``x'' can be calculated as:
\begin{equation}
\label{eq:mass_change}
\dot{m}_x = - 4 \pi R^{2} P_{x} (T_{pl}) \sqrt{\frac{\mu_{x}}{2 \pi R_g T_{pl}}} ,
\end{equation}
where $P_{x}$ is the saturated vapor pressure \footnote{The term saturated vapor pressure represents the partial pressure measured in laboratory experiments, serving as a proxy for the sublimation rate of a
chemical compound (see \citet{2009F} for more details)} of the chemical compound x, $T_{pl}$ is the  planetesimal surface temperature (it will be presented later in this section),  $\mu_{x}$ is the molecular weight of
the element x, and $R_g$ is the ideal gas constant.

Planetesimals in reality are made of a mixture of different materials such as irons, silicates, ices and organics. An accurate determination of the mass ablation rate of planetesimals phases requires sophisticated radiation-hydrodynamic and computational fluid dynamics simulations and laboratory experiments~\cite[e.g.][]{podolaketal88,makinde13}. In this work, we follow previous studies and, for simplicity, we assume that planetesimals are made of either pure water-ice rich phase, ``silicate'', or ``organics'' chemical compounds~\citep{mosespoppe17,2021A&A...648A.112E}. Each of these scenarios envision that a respective materials are the major phases of the planetesimal bulk composition and all volatile compounds are carried by them. It also implies that if gas-frictional heating leads to sublimation of any specific major phase, less abundant phases well mixed with the major phases will be also ablated and recycled into the gaseous disk~\citep[e.g.][]{2021A&A...648A.112E}.  We have performed simulations assuming four arbitrary distinct compositions for our planetesimals, namely i) water-ice rich (${\rm H_2O}$), which may be a good proxy to represent planetesimals formed beyond the water snowline~\citep{mosespoppe17,2021A&A...648A.112E}; ii) water-ice poor planetesimals, made of mineral enstatite (${\rm \rm MgSiO_3}$;  \cite{CAMERON1985285}), which may consistent with planetesimals formed inside the water snowline; iii) organic-rich , made of benzo(a)pyrene (${\rm C_{20}H_{12}}$; \cite{mosespoppe17}), which may be consistent with planetesimals formed beyond the tar-line where carbon can condense; and and fayalite-rich planetesimals (${\rm Fe_2SiO_4}$; \cite{nagaharaetal94}), a mineral found in carbonacenous and ordinary chondrites. By testing the effects of ablation on planetesimals with these different phase-specific compositions, we qualitatively cover a large range of the spectrum of possible planetesimals' compositions.  Next, we describe how we calculate the planetesimal surface temperature used in Eq. \ref{eq:mass_change}.

\subsubsection{Planetesimal Surface temperature}

The mass ablation rate of a planetesimal depends on its surface temperature. We assume that the surface temperature of a planetesimal is in equilibrium with the local gas-disk temperature, frictional heating with the gas, and cooling due to sublimation \citep{refId0}. The equilibrium surface temperature of a planetesimal in a protoplanetary disk can be calculated as \citep[e.g.,][]{1924MWRv...52..488W, 1971NASSP.252..219B,2004A&A...418..751C,refId0,2021A&A...648A.112E}

\begin{equation}
\label{eq:temp}
T^{4}_{pl} = T^{4} +  \frac{C_d \rho v_{rel}^{3}}{32 \sigma_{sb} } - \frac{P_{x}(T_{pl})}{\sigma_{sb}} \sqrt{\frac{\mu_{x}}{8 \pi R_g T_{pl}}} L_{x}
\end{equation}
The first term in Equation \ref{eq:temp} represents the gas temperature at the position of the planetesimal, while the second term arises from the heating due to  gas-planetesimals aerodynamic friction. The last term represents the cooling due to the release of latent heat during ablation of an element x. The Stefan-Boltzmann constant is represented by $\sigma_{sb}$ and the latent heat of  vaporization of the element x is given by $L_{x}$. The latent heat of vaporization and  saturated vapor pressures depend on the planetesimal composition as described in the next sections. It is important to note that the cooling rate also depends on the planetesimal surface temperature, therefore, in order to solve Eq. \ref{eq:temp}, we invoked the bisection method.
\subsubsection{Water-ice rich planetesimals}

We envision that a water-ice rich composition is a good proxy for undifferentiated planetesimals formed beyond the water snowline in the disk \citep{refId0,2021A&A...648A.112E} that experience none or limited water-loss due to internal heating~\citep{monteuxetal18,newcombeetal23}. A water-ice rich composition could also be relevant for large planetesimals that underwent ice-rock differentiation, such as Europa and Ganymede, if these objects existed in the heliocentric planetesimal population.
We model surface ablation of water-ice rich planetesimals by assuming a bulk density, molecular weight, and latent heat of vaporization of pure water, as given in Table \ref{tab:waterprop}.
We calculate the saturated vapor pressure using the fitting coefficients provided by \citet{2009F}  (see Table \ref{tab:waterprop}) using the formulas:

\begin{equation}
\label{eq:saturated}
\ln\left( \frac{P_{\text{H}_2\text{O}}(T)}{P_t} \right) = \frac{3}{2} \ln\left(  \frac{T}{T_t} \right) + \left(1 - \frac{T_t}{T} \right) \eta\left(\frac{T}{T_t}\right)
\end{equation}

\begin{equation}
\label{eq:saturated2}
\eta\left(\frac{T}{T_t}\right) = \sum_{i=0}^{6} \alpha_i \left(  \frac{T}{T_t} \right)^i
\end{equation}
where $P_t$ and $T_t$ stand for the temperature and pressure at the triple point, respectively. We refer the reader to \citet{2009F} and \citet{refId0} for further details.

\begin{table}
	\centering
	\caption{The molecular weight ($\mu_{H_{2}O}$) and the latent heat of vaporization of the ice-water ($L_{H_{2}O}$) and the interpolation coefficients $\alpha_i$ of saturated vapor pressure of ice water.
	According \citet{2009F} these coefficients are only accurate for the ice water in the range of temperatures between 0 and 273.16 K. The data are sourced from \citet{podolaketal88}.}
	\label{tab:waterprop}
\begin{tabular}{ccc}
		\hline
        $\mu_{H_{2}O}$ (Kg/mol) & $0.0180153$ &  \\
		$L_{H_{2}O}$ (J/Kg) & $2.80 \times 10 ^{6}$ & \\
		$\alpha_0$ & $20.9969665107897$ & \\
		$\alpha_1$ & $3.72437478271362$ & \\
		$\alpha_2$ & $-13.9205483215524$ &  \\
		$\alpha_3$ & $29.6988765013566$  & \\
		$\alpha_4$ & $-40.1972392635944$ &  \\
		$\alpha_5$ & $29.7880481050215$ & \\
		$\alpha_6$ & $-9.13050963547721$ & \\
		\hline
	\end{tabular}

	\centering

\end{table}

\subsubsection{Water-ice poor planetesimals}

We assume that our water-ice poor planetesimals are rich in mineral enstatite. Enstatite is common in some classes of achondrite (differentiated) meteorites and are abundant minerals in enstatite chondrite meteorites. Enstatite chondrites are most likely to have formed in the inner solar system rather than beyond the snowline \citep{2018SSRv..214...36A}.
Consequently, this composition represents an extreme end-member scenario among our selected planetesimal compositions. Following \citet{CAMERON1985285},
we calculate the mass ablation rate for water-ice poor planetesimals assuming the following
parameters: $L_{s} =5 \times 10 ^{6}$ (J/Kg), $\mu_{s} = 0.100387$ (Kg/mol), and $\log{P}s = 13.176 - 24605.5/T{pl}$, where ${P}_s$ is given in $dyne/cm^{2}$ \citep{CAMERON1985285}.


\subsubsection{Fayalite-rich and organic-rich planetesimals}

The melting of the ice component of water-rich planetesimals caused by internal heating and differentiation~\citep[e.g.][]{2011EP&S...63.1193W,2012A&A...543A.141N} followed by chemical reactions between solid and fluid leads to aqueous alteration and the formation of secondary minerals \citep{GRIMM1989244}. Several meteorite classes show evidence of aqueous alteration resulting in hydrous oxygen-bearing secondary minerals that include phyllosilicates, magnetite, sulfides, carbonates,
ferromagnesian olivines,  and others \citep{krot05,2015NatCo...6.7444D}.
Fayalite, the Fe-end member of olivine, is found  both in the matrix and chondrules of carbonacenous chondrites~\citep{krot05} and is the outcome of aqueous alteration in the parent body~\citep{2015NatCo...6.7444D}. Fayalite is a common planetary material~\citep{2021ApJ...913L..31S}. We test the effects of surface ablation for planetesimals rich in fayalite. The parameters chosen for planetesimals rich in fayalite were $L_s = 5 \times 10^6$ (J/kg), $\mu_s = 0.100387 (kg/mol)$, and $\log{P}s = 199 - 60409/T{pl}$ (given in bar; see \citet{NAGAHARA19941951}).

The organic content of CC meterorites is high in relation to total carbon content~\cite[e.g.][]{pizzarelloetal06} and mainly consists of an insoluble kerogen-like material. For completeness, we also performed simulations considering planetesimals with an organic-rich composition. For these planetesimals we adopted $L_s = 1.5 \times 10^6$ (J/kg), $\mu_s = 0.077 (kg/mol)$, and $\log{P}s = 9.110 - 7100/T{pl}$, where ${P}s$ is given in atm. These parameters  are appropriate for benzo(a)pyrene ($C_{20}H_{12}$; see \citet{2017Icar..297...33M}), an organic compound found in carbonaceous chondrites.

    \begin{figure}
 \includegraphics[width=\columnwidth]{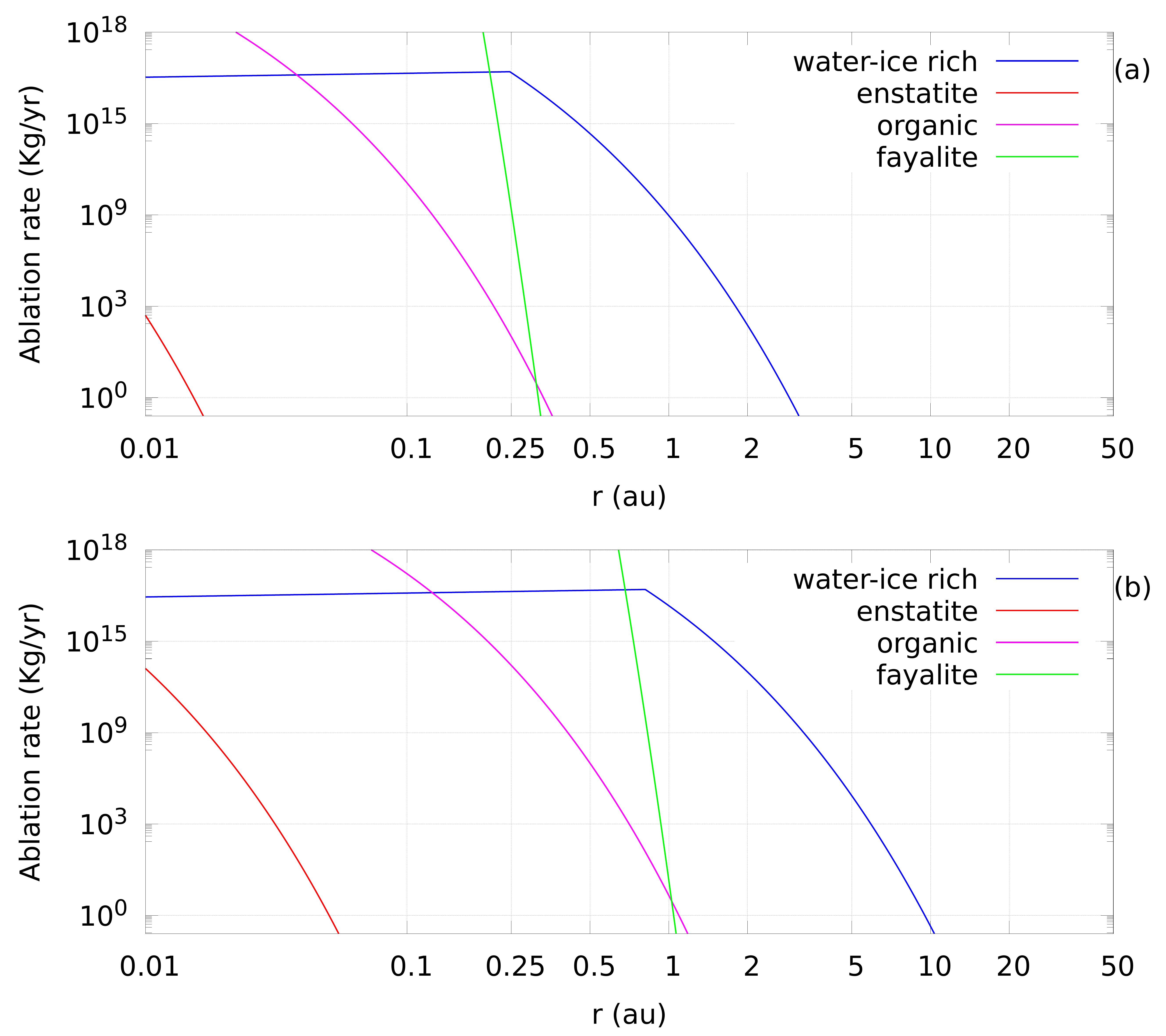}

          \caption{The ablation rate of water-ice rich (blue line), enstatite (red line), organics (magenta line) and fayalite (green line) for
          a 100-km-sized planetesimal as a function of the heliocentric distance and considering a \textit{cold disk} (i.e, snowline at 1 au) (Panel a) and a \textit{hot disk}  (i.e, snowline at 3.3 au) (Panel b). In this experiment, we neglect the effects of heating due to gas friction (i.e we impose $v_{rel}=0$). The polynomial fit representing the water-ice saturated pressure tends to minus infinity  as the temperature decreases, we follow \citet{2021A&A...648A.112E}
          and introduce a floor-value at 60 k to prevent numerical issues. \label{fig:proporties}}
\end{figure}

\subsubsection{Mass ablation rates of planetesimals with different composition }

Before starting our simulations we illustrate the effects of surface mass ablation using an idealized scenario. Figure \ref{fig:proporties} shows the mass ablation rate of 100-km  sized planetesimals as a function of heliocentric distance for  different planetesimal compositions. In this experiment, we neglect the effects of heating due to gas friction (i.e we impose $v_{rel}=0$), but we account for local thermal heating from the gas and  surface cooling (see Eq. 6). We are mainly interested in demonstrating how  the effects of the ablation caused by the local gas temperature change depending on the planetesimal composition and, finally, show that our calculations are broadly consistent with those of previous studies (see \citet{2021A&A...648A.112E}). Figure \ref{fig:proporties}-(a) and (b) show mass ablation rates of planetesimals in our ``cold'' and ``hot'' disks, respectively. These figures show that water-ice rich planetesimals are relatively ``fragile'' and start to lose significant mass via ablation if reaching  distances as close as 2.5-10 au from the Sun. On the other hand, planetesimals with enstatite-like compositions are relatively more robust and are only ablated if crossing the innermost regions of the disk ($<$0.1~au). Notice that ablation starts at a temperature well below the nominal condensation temperature of the corresponding species. The 'knee' in Figure \ref{fig:proporties} is attributed to the polynomial fit representing the water-ice saturated pressure. This saturated pressure tends toward minus infinity as the temperature decreases. Following the approach of \citet{2021A&A...648A.112E}, we have introduced a floor value at 60 K to prevent numerical issues.

\section{Numerical Simulations}\label{Sec:Numerical_Simulations}

\subsection{Numerical step}

We used the REBOUND code \citep{Rein2012, ReinSpiegel2015, ReinTamayo2015} to perform N-body simulations. We model the evolution of planetesimals evolving under the gravitational influence of the Sun, Jupiter, and Saturn. Planetesimals also experience the effects of gas-drag due to an underlying gaseous protoplanetary disk (see Section \ref{gas}). Planetesimals do not
gravitationally interact among themselves.

Our simulations were performed using the integrator IAS15 \citep{ReinSpiegel2015} and an initial timestep of 0.1 years. Collisions between planets and planetesimals are modelled as perfect merging events \citep{2012ApJ...745...79L,10.1093/mnras/stab3203}, conserving mass and linear momentum.  Planetesimals are considered ejected from the simulation if the heliocentric distance gets larger than 1000 au or orbital eccentricity gets larger than 1. We do not model the accretion of terrestrial planets in these simulations, but only the implantation of planetesimals in the asteroid belt.

We model the effects of surface ablation by solving Eq. (\ref{eq:temp})  and  calculating the mass loss of individual planetesimals due to their surface ablation at each timestep (Eq. (\ref{eq:mass_change})). We consistently adjust the mass of each planetesimal that suffered ablation and update its size while assuming that its initial bulk density remains constant. We keep track of the amount of mass ablated from the planetesimals at each timestep. If a planetesimal loses 100\%  of its initial mass, it is removed from the simulation.

\subsection{Natal regions of the planetesimals}

Planetesimals are initially radially distributed in the disk in two regions, defined as Region 1 and Region 2.  In our GT simulations, planetesimals in Region 1 were initially distributed between the orbits of Jupiter and Saturn (3.6--4.4 au), while those in Region 2 were initially
distributed beyond the orbits of the giant planets, up to 13 au.
In our insituJSrapid simulations, planetesimals in Region 1 were initially distributed between the orbits of Jupiter and Saturn (5.0--6.55 au) and beyond Saturn's orbit, up to 15 au. Finally, in our migJSrapid simulations, planetesimals in Region 1 were initially distributed from 5--7 au and those in Region 2 were initially distributed from 7--20 au. Planetesimals are randomly distributed between the radial limits of Region 1 and Region 2.

\subsection{The initial size distribution of planetesimals}
\label{sec:initialsize}
We performed simulations considering two different initial size distribution of planetesimals.  In the first one, hereafter defined as NO-SFD simulation, we assumed that the entire initial planetesimal population has a single given size (diameter). For water-ice rich and water-ice poor planetesimals, we performed simulations considering planetesimal populations consistent of planetesimals with diameters  of $1$ km, $10$ km, $100$ km, and $1000$ km. Our simulations start with at least 5000 planetesimals, but in the case of 100 km-sized planetesimals we have performed high resolution  simulations where in the initial number of planetesimals is set to as 100,000. For organic and fayalite rich planetesimals, we only performed simulations with populations of 100-km sized planetesimals.

In the second set of simulations (hereafter refereed to as SFD simulation), we assumed an initial size-frequency distribution (SFD) for the planetesimal population. As these simulations are computationally expensive, in order to save computational time, we only performed simulations assuming an initial SFD for the water-ice-rich and organic planetesimals cases.
Our initial planetesimal populations were created following a differential distribution with $dN/dD \propto D^{-p}$, with $p=5$ and assuming a diameter binning of 10 km in the range of 100 km $\leq D \leq$ 1000 km. We choose a differential distribution for simplicity. Nevertheless, the power slope ($p$) of the differential distribution relates to the power slope ($q$) of the cumulative distribution  as $p=q$ + 1 \citep[e.g.][]{marschalletal22}. This implies the power slope for our cumulative SFD is $q=4$ in $N(>D) \propto D^{-q}$.
The motivation behind our chosen SFD comes from recent  planetesimal formation models \citep{2005YoudinGoodman,simonetal16}. It is also consistent with the hypothesis that planetesimals are born with  a typical size of 100 km or larger \citep{2009Icar..204..558M, 2020ApJ...901...54K, 2021ApJ...911....9K}.

When creating our initial planetesimal population from our assumed SFD, the largest  planetesimal size was set to as $D=1000$ km, and the smallest one was set to as  $D=100$ km ({\color{blue} Deienno et al. 2022}). Our simulations assuming a  planetesimals population with an initial SFD started with 1.831 million planetesimals in each region (3.662 million planetesimals total in Region 1 and Region 2). We have only performed simulations with an initial SFD for the  \textit{cold disk} scenario (i.e. the snowline is initially at 1 au), also because of the high computational cost of these simulations.  In all our simulations, the initial eccentricities and orbital inclinations of all planetesimals are set to as $10^{-3}$. Other orbital angles are randomly selected between $0$ and $360$ degrees. In the next section, we present the results of all our numerical simulations.

\section{Results}\label{Sec:Results}

\subsection{NO-SFD simulation set}
We start by presenting the results of our NO-SFD simulation set following our \textit{cold disk} scenario (See Section \ref{sec:initialsize}).

\subsubsection{Dynamical evolution}
\label{sec:devol}

\begin{figure*}
 \includegraphics[scale=0.4]{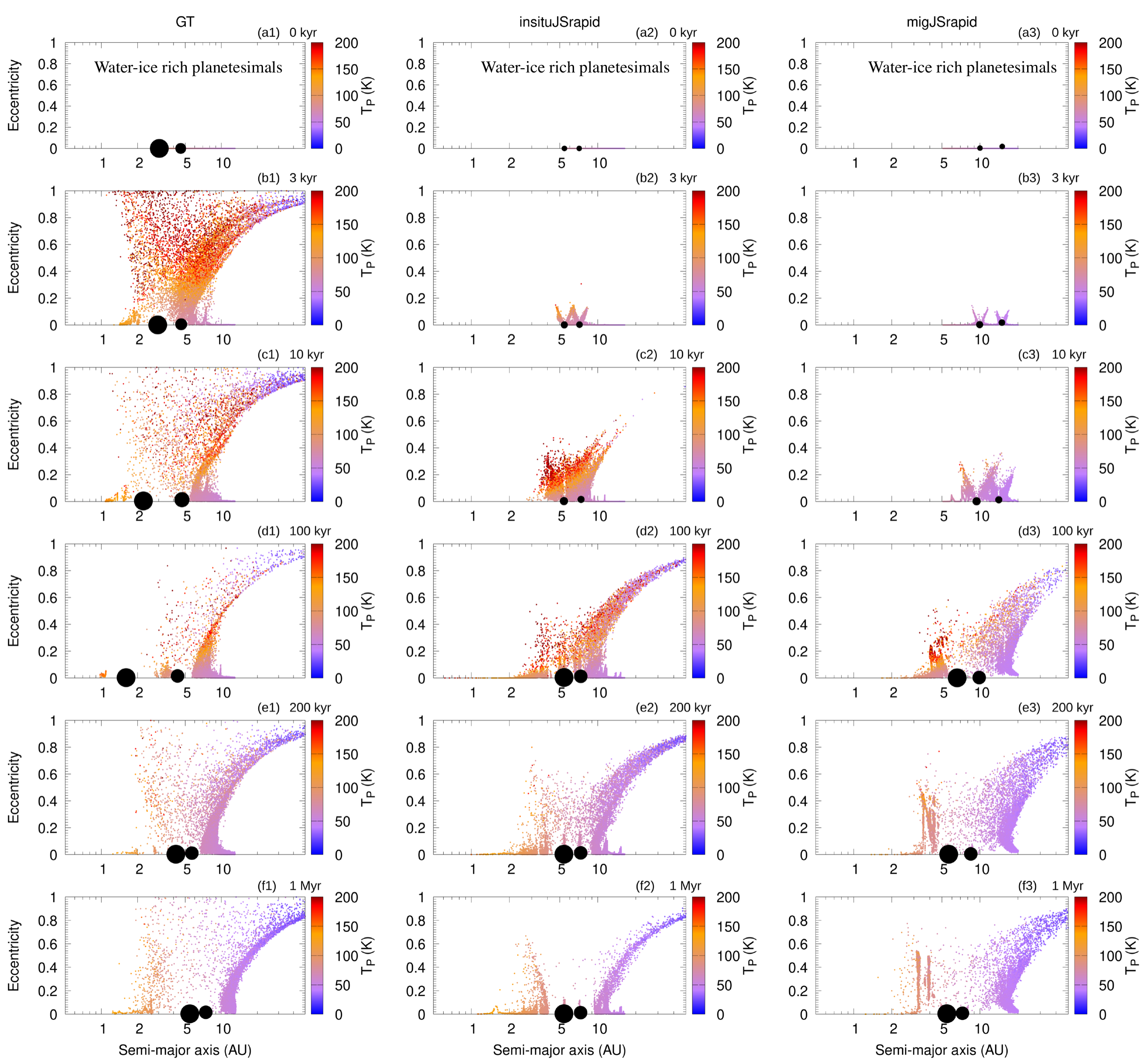}
          \caption{Eccentricity/Semi-major axis/temperature plot portraying the dynamical evolution of the cases GT (Panels (a1)--(f1)), insituJSrapid (Panels (a2)--(f2)) and migJSrapid--Sim2 (Panels (a3)--(f3)). Jupiter and Saturn and their relative sizes are represented by black-filled circles. The color box represents the temperature of the planetesimals. The instant where the snapshot was taken with respect to the simulation time is shown at the top of each panel. We used the diameters  of $1$ km, $10$ km, $100$ km, and $1000$ km for the planetesimals.}
\label{fig:case_1_sim_2_a_e}
\end{figure*}

Figure \ref{fig:case_1_sim_2_a_e} shows snapshots of the semimajor axis, eccentricity and temperature of  water-ice rich planetesimals for simulations considering different migration histories of Jupiter and Saturn: GT (Panels (a1)--(f1)), insituJSrapid (Panels (a2)--(f2)), and migJSrapid (Panels (a3)--(f3)). During the growth and dynamical evolution of the giant planets, planetesimals close to Jupiter and Saturn are scattered to the inner and outer parts of the Solar System. Most planetesimals in Region 1 (note that the definition of Region 1 and 2 changes depending on the migration history of Jupiter and Saturn) do not remain in their initial birth locations (as also observed in \citet{2021A&A...648A.112E}) and ultimately end up scattered by the planets. Scattered planetesimals reach moderate and high eccentricity orbits. High eccentricities lead to higher relative velocities with the gas, resulting in increased planetesimal high surface temperatures and mass loss rates due to ablation.

As expected, Figure \ref{fig:case_1_sim_2_a_e} shows that planetesimal tend to reach high surface temperatures when they are scattered into eccentric orbits and come closer to the Sun. Water-ice rich planetesimals that are scattered towards the inner solar system may reach surface temperatures exceeding $\sim$100 K.  At these temperatures water-ice  planetesimals start to lose mass via ablation efficiently, shrinking in size and/or experiences complete destruction by ablation. As a result, there are no water-rich planetesimals remaining in the innermost part of the disk ($a<1$ au) by the end of the gas disk phase. This is not necessarily the case for planetesimals of other compositions. The disruption of CC planetesimals for $q<1$ au has important implications in the terrestrial planet region. Isotopes (more details in \citet{SAVAGE2022115172}) reveal that about 30\% to 50\% of Earth's volatile budget is inherited from CC planetesimals. Therefore, the complete destruction of CC planetesimals by ablation may be problematic in explaining the source of volatile elements for Earth.

The dynamical evolution of planetesimals with water-ice poor, organic-rich, and faylite-rich compositions are qualitatively similar to those of illustrated Figure \ref{fig:case_1_sim_2_a_e}. However, it is worth pointing out that water-ice poor planetesimals are only significantly affected by surface ablation if their surface temperatures becomes higher than $\sim$800 K. These temperatures are only reached in the innermost parts of the gaseous disk (e.g. when eccentricities are greater than 0.8). As a result, the majority of water-ice poor planetesimals that are scattered within Jupiter's orbit -- regardless of the migration histories of Jupiter and Saturn --  tend to survive their injection phase into the inner system, unless they get as close as 0.1~au from the Sun, where ablation of enstatite-rich materials becomes significantly efficient. In the next section, we will further discuss on the implantation of planetesimals into the asteroid belt of different compositions and initial sizes.

\subsubsection{Implantation of planetesimals into the asteroid belt}
\label{plaaftergasasteroidbelt}

\begin{figure*}
 \includegraphics[scale=0.27]{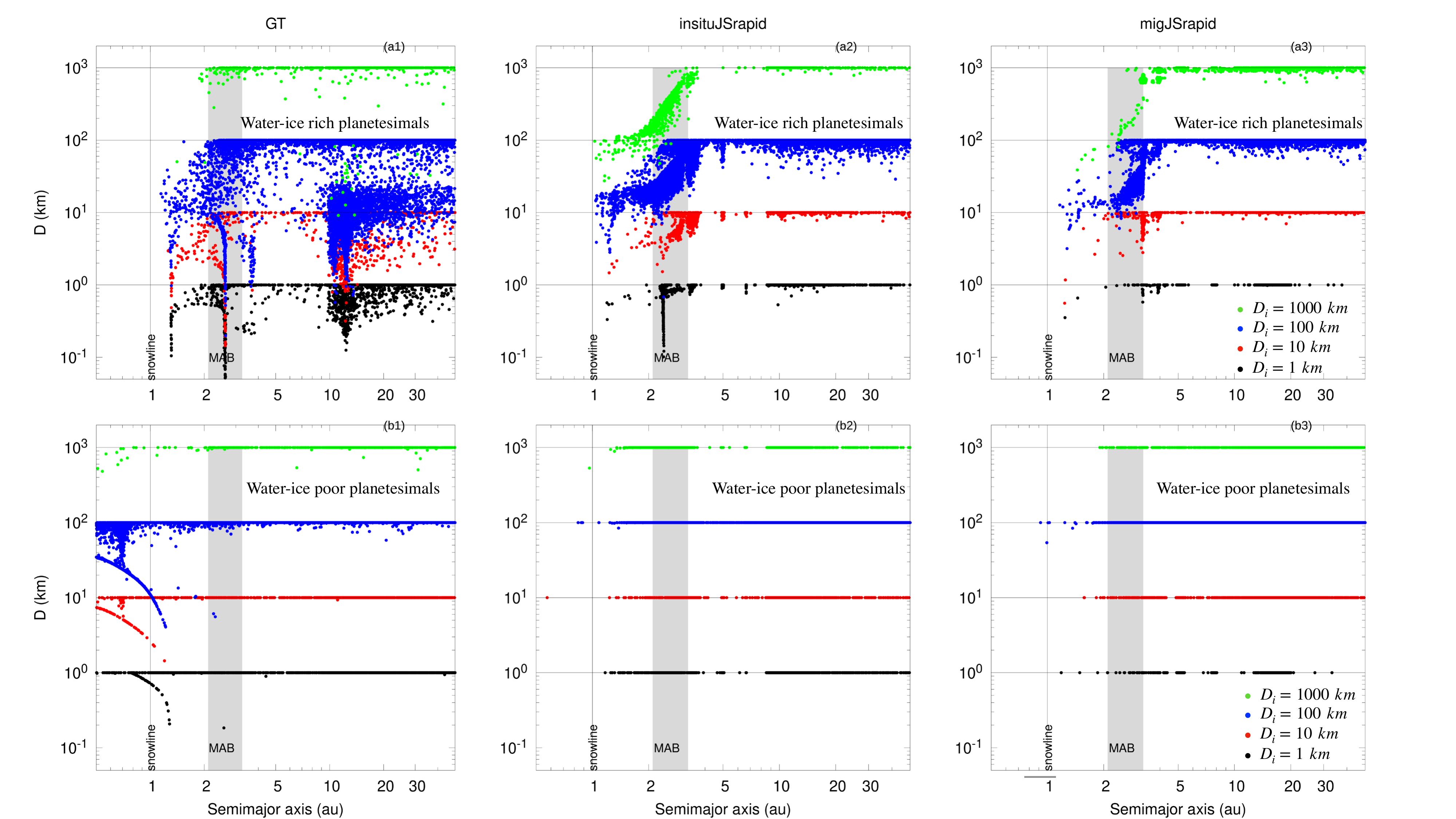}

          \caption{Final diameter distributions of planetesimals (color coded) from the simulations cases GT (Panels (a1)--(b1)), insituJSrapid (Panels (a2)--(b2)) and
          migJSrapid (Panels (a3)--(b3)) for water-ice rich (top panels) and poor planetesimals (bottom panels). The asteroid belt -- defined by the orbital region with $2.1 < a < 3.25$ au, $e < 0.4$ and $i < 40$ degrees-- is shaded. The vertical line represents the location of the water snowline. Horizontal lines and black, red, blue and green points are a reference for D = 1 km, 10 km, 100 km, and 1000 km, respectively (the initial size of planetesimals in our sample). \label{fig:water_ice_1}}
\end{figure*}

Figure \ref{fig:water_ice_1} shows the diameter distributions of water-ice-rich planetesimals (Panels (a1)--(a3)) and water-ice-poor planetesimals (Panels (b1)--(b3)) at the end of the disk phase for all giant planet migration histories considered in our simulation: GT, insituJSrapid, and migJSrapid scenarios. The shaded region -- defined by the orbital region with $2.1 < a < 3.25$ au, $e < 0.4$ and $i < 40$ degrees-- in Figure \ref{fig:water_ice_1} corresponds to the asteroid belt region. Again, this analysis corresponds to the results of our NO-SFD simulation set following our cold disk scenario.
In Figure \ref{fig:water_ice_1}, each panel shows the final diameter distribution of planetesimals produced from four different simulations starting with  different initial planetesimal diameters (1, 10, 100 and 1000~km). The top panels shows the results of simulations considering water-ice rich planetesimals (Panels (a1)--(a3) of Fig. \ref{fig:water_ice_1}). The bottom panels (Panels (b1)--(b3) of Fig. \ref{fig:water_ice_1}) shows the results of simulations consider water-ice poor planetesimals.

 Water-ice rich planetesimals are efficiently implanted into the asteroid belt but with reduced diameters relative to the initial ones (compare planetesimal sizes within the gray region with their respective original diameters, i.e., horizontal lines). The horizontal lines at D = 1, 10, 100 and 1000~km are  added as a guide to illustrate the four initial diameters of planetesimals in our simulations. Planetesimals plotted slightly below a given horizontal line (or between  horizontal lines) experienced ablation and had their diameter reduced. As one can see in all top panels, a significant fraction of our water-ice rich planetesimals, regardless of initial diameter, experienced ablation during implantation into the asteroid belt. It is interesting to note that all migration histories did not implant water-ice rich planetesimals inside 1~au. This either may be an outcome of surface ablation ``destroying'' planetesimals  reaching the innermost regions of the disk or may be simply reflect inefficient delivery of planetesimals inside 1~au. We will address this question by comparing these results with those of water-ice poor planetesimals.

Unlike our simulations with water-ice rich planetesimals, simulations considering water-ice poor planetesimals (Panels (b1)--(b3) of Figure \ref{fig:water_ice_1}) show that the final sizes of water-ice-poor  {\it implanted} planetesimals in the asteroid belt remain nearly unchanged from their initial sizes, regardless of the giant planets' migration histories. This scenario essentially recovers the outcome of classical studies neglecting the effects of surface ablation ~\citep{Walshetal2011, raymond17}.
Simulations considering the insituJSrapid and migJSrapid scenarios implanted very few planetesimal inside 1~au at the end of the gas disk phase, typically 1 or 2 objects. It is clear from that observation that the insituJSrapid and migJSrapid scenarios tend to deliver much less material inside 1~au than the GT case, where many planetesimals survived inside 1~au (Figure \ref{fig:water_ice_1})-b1). By comparing panels (a1) and (b1) of Figure \ref{fig:water_ice_1}, we can conclude that the lack of water-ice-rich planetesimals inside 1~au in Figure \ref{fig:water_ice_1}-(a1) is a consequence of ablation destroying water-rich planetesimals when get they closer than 1~au from the Sun. As Jupiter migrates inwards down to 1.5~au in the GT evolution, planetesimals are pushed inwards reaching eccentricities as high as $\sim$0.9, which enhances the efficiency of surface ablation relative to the insituJSrapid and migJSrapid evolutions.
In the GT evolution (Figure \ref{fig:water_ice_1})-(b1)), even water-ice poor planetesimals initially larger than $\sim$10-100~Km were significantly ablated and pushed at distances as close as 0.5 au from the Sun. For a planetesimal on a high eccentricity orbit frictional heating can significant boost the effects of surface ablation. In our GT simulation, enstatite-like planetesimals at $\sim$1~au were also ablated which is different from that ideal case of Figure \ref{fig:proporties}, where $v_{\rm rel}$ was ideally assumed equal to zero.

\begin{figure*}
 \includegraphics[scale=0.25]{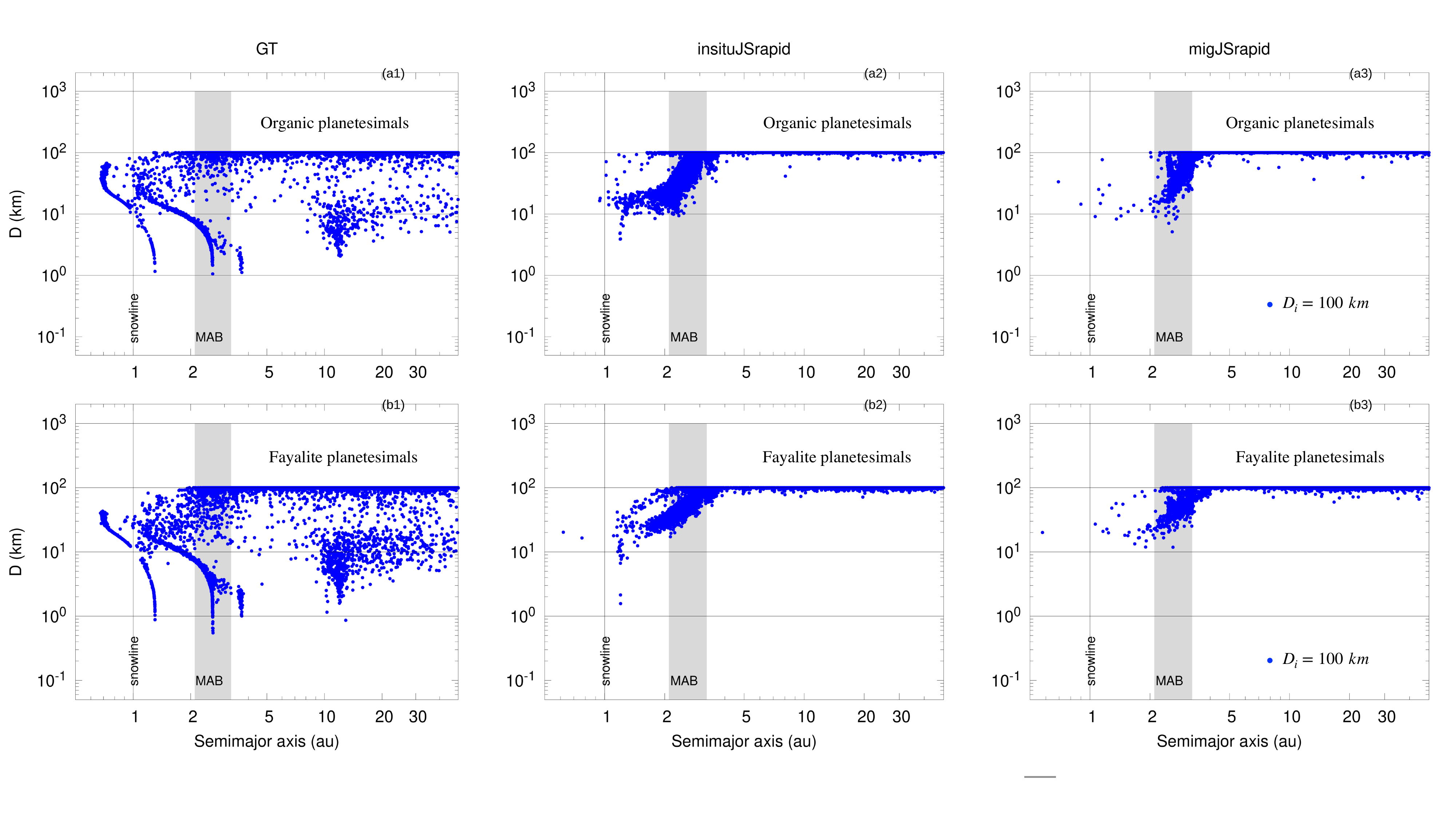}

          \caption{The same of Figure \ref{fig:water_ice_1} but for organic and fayalite planetesimals \label{fig:org_fay_1}}
\end{figure*}

Figure \ref{fig:org_fay_1} shows the final diameter distributions of organic (Panels (a1)--(a3)) and fayalite-rich (Panels (b1)--(b3)) planetesimals from simulations following the GT, insituJSrapid, and migJSrapid evolutions.
As explained before (Section \ref{sec:initialsize}), for these two planetesimal compositions we only performed simulations considering planetesimals initially of D=100~km.  Overall, our results show that both  organic and fayalite rich planetesimals implanted in to the asteroid belt tend to experience surface ablation and shrink in size during the gas-assisted implantation process. However, water-ice rich planetesimals tend to experience relatively more ablation than organic and fayalite rich planetesimals, as also suggested by Figure \ref{fig:proporties}.

Altogether, our results suggests that unless, planetesimals surface materials were as resistant to aerodynamic and thermal heating as  enstatite-like minerals mass surface ablation may have significantly reduced the sizes of planetesimals during gas-assisted implantation into the asteroid belt. In the next section, we perform statistical analysis of how surface ablation changes the initial size-distribution of planetesimals implanted into the asteroid belt.


\subsubsection{Size frequency distribution and statistics on planetesimal implantation}
\label{organic_faylite}

    \begin{figure*}
 \includegraphics[scale=0.18]{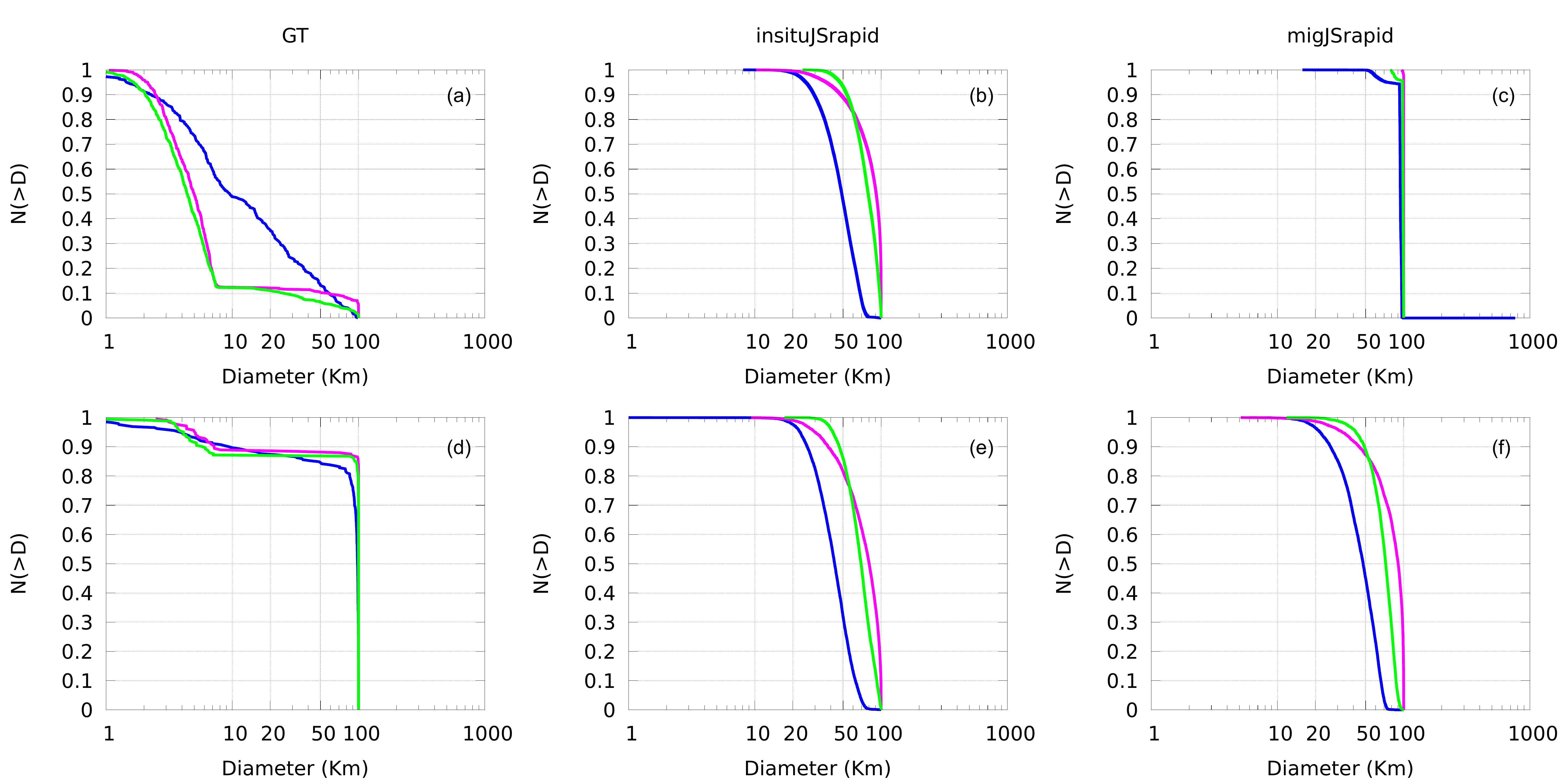}

          \caption{Panels (a) to (c) show the cumulative normalized sizes of water-ice-rich (blue line), organic (magenta line), and fayalite planetesimals (green line), respectively, initially with a size of 100 km, implanted in the Asteroid Belt from Region 1 at the end of the gas disk phase in simulation sets GT, insituJSrapid, and migJSrapid. Panels (d) to (f) show the cumulative normalized sizes of these planetesimals, also implanted in the asteroid belt, but from Region 2. \label{fig:cumativesize_all}}
\end{figure*}

In Figure \ref{fig:cumativesize_all}, we show that regardless of the migration history of the giant planets and planetesimals' compositions, at least a fraction of the implanted planetesimals is affected by surface ablation. Figure \ref{fig:cumativesize_all} specifically shows the cumulative normalized size distributions of water-ice-rich (blue line), organic (magenta line), and fayalite rich planetesimals (green line) implanted into the asteroid belt. These results correspond to those shown in  Figures \ref{fig:water_ice_1} and \ref{fig:org_fay_1}, where planetesimals were assumed to have  an initial size of D=100 km. The top panels of
Figure \ref{fig:cumativesize_all} show the distribution of asteroids implanted from  Region 1 at the end of the gas disk phase in simulations with different migration histories of the giant planets.  The bottom panels of Figure \ref{fig:cumativesize_all} show the corresponding distributions for Region 2. We present  our analysis dividing the planetesimal populations between their original source locations, namely Region 1 and Region 2, in order to demonstrate how ablation may affect planetesimals in each of these regions.

The very migration history of the giant planets results in different levels of surface ablation (affecting the implanted population). For instance,  in the GT scenario, only $\sim$8-10\% of the initial 100-km water-ice-rich, organic, and fayalite planetesimals from Region 1 implanted into the asteroid belt have sizes larger than 50 km (Fig. \ref{fig:cumativesize_all} (a)). In contrast, in the insituJSrapid model, about 50\% of the initial 100-km water-ice-rich planetesimals from Region 1 are implanted in the Asteroid Belt with sizes larger than 50-70 km. Surface ablation also impacts Region 1 and Region 2 in different ways.
In the GT model, surface ablation shows a weaker impact on implanted planetesimals from  Region 2, compared to Region 1 (see panels (a) and (b) of Figure \ref{fig:cumativesize_all}). In the insituJSrapid case, most implanted planetesimals, regardless of the composition and original Region, have sizes between 20 and 100~Km, showing a similar impact of surface ablation for implanted planetesimals of  Region 1 and 2. In the migJSrapid scenario, planetesimals implanted from Region 1 suffer limited size reduction due to ablation but those implanted from  Region 2 have experienced a more strong effect.

Note that the normalized implanted distributions of Figure \ref{fig:cumativesize_all} do not show the absolute number of planetesimals implanted into the asteroid belt and the efficiency of ablation because planetesimals may have been destroyed by ablation during the implantation process. In other words, the total number of planetesimals implanted in the asteroid belt in each of these simulations is different so this figure can not be used to claim that the effect of ablation is weaker/stronger for Region 1/2. It can only be used to compare the impact of ablation on the final belt-implanted planetesimal populations. The fraction of the initial 100-km sized planetesimals implanted in the asteroid belt for any history and composition of the planetesimals is shown in Table \ref{tab:fraction_ab}.

 \begin{table}
	\centering
	\caption{Fraction of planetesimals implanted into the asteroid belt in simulations where the planetesimal initial size is $D=100$ km, based on an initial total of $1 \times 10^5$ objects. Different compositions of planetesimals are considered. We show the total final fraction of implanted planetesimals regardless of the planetesimal final size.}
	\label{tab:fraction_ab}
	\begin{tabular}{ccccc}
		\hline
		Composition: & Water-ice rich & Organic & Fayalite \\
		\hline
		GT: Region 1 &  $3.2 \times 10^{-3}$ & $5.0 \times 10^{-3}$  & $8.6 \times 10^{-3}$  \\
		insituJSrapid:  Region 1  &  $2.8 \times 10^{-1}$  &  $2.7 \times 10^{-1}$  & $3.5 \times 10^{-1}$  \\
		migJSrapid: Region 1  &  $2.0 \times 10^{-2}$   & $1.1 \times 10^{-2}$   & $1.2 \times 10^{-2}$\\
		GT: Region 2 &  $2.9 \times 10^{-3}$  & $2.1 \times 10^{-3}$  & $2.4 \times 10^{-3}$   \\
		insituJSrapid:  Region 2  &  $6.0 \times 10^{-2}$ & $5.7 \times 10^{-2}$  & $6.7 \times 10^{-2}$  \\
		migJSrapid: Region 2  &  $6.9 \times 10^{-2}$   & $4.5 \times 10^{-2}$   & $5.1 \times 10^{-2}$ \\
		\hline
	\end{tabular}
\end{table}


Our simulation shows that most planetesimals implanted in the GT scenario are small, with typical sizes of 5~km. In the insituJSrapid scenario, most planetesimals have sizes of about 50~km or larger.





\subsubsection{The effect of the initial temperature of the disk}
\label{snowline}

It is legitimate to wonder how our results would change if the protoplanetary disk was hotter than what we have considered in our nominal simulations. For completeness, we now present the results of  simulations where the disk water snowline is at the outer edge of the main asteroid belt (See Section \ref{temperature} for details).

\begin{figure*}
 \includegraphics[scale=0.26]{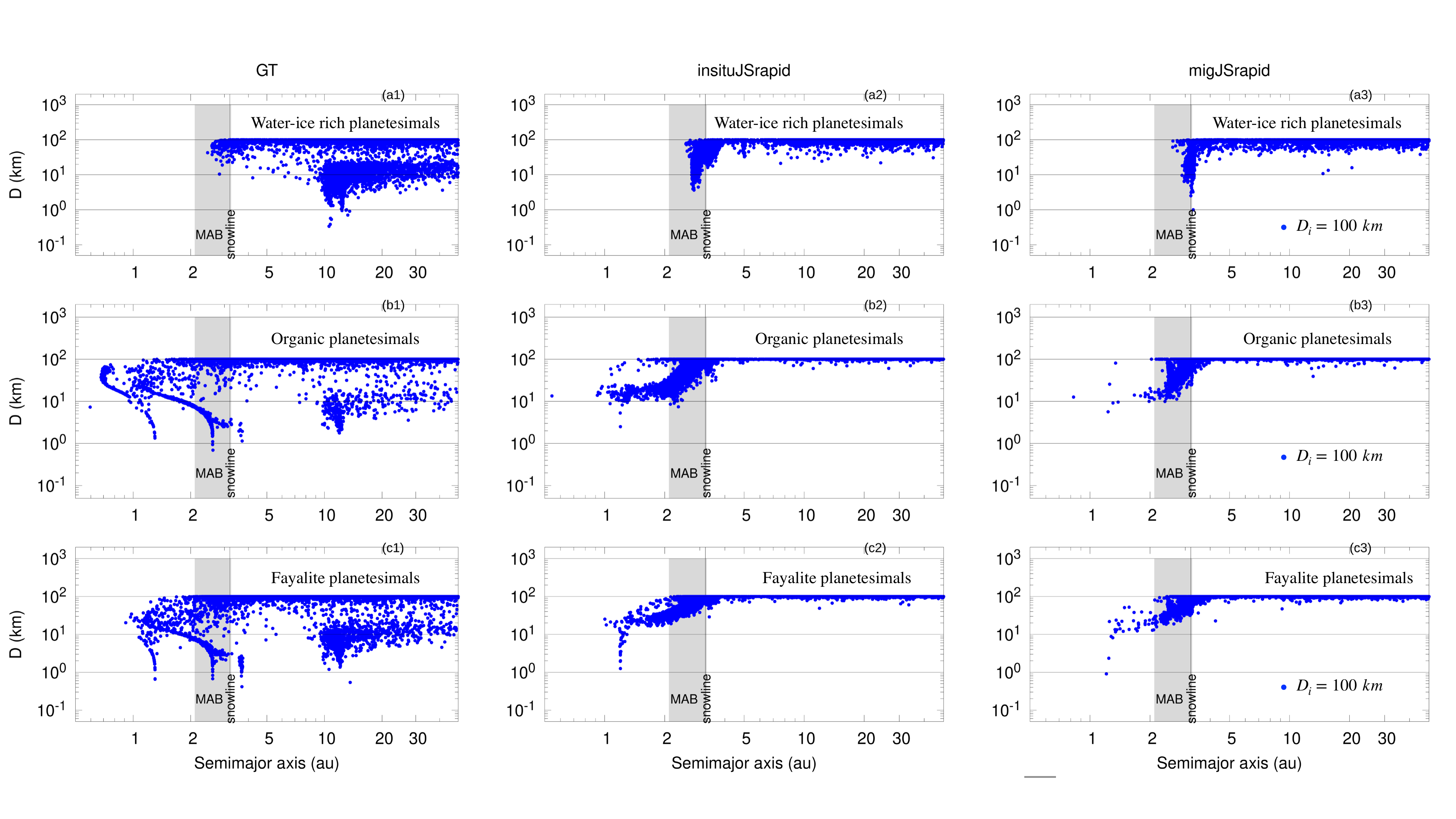}
          \caption{The same of Fig. \ref{fig:org_fay_1} but for a protoplanetary disk with temperature of  $T = 250 K (r/au)^{-3/7}$ (hot disk, i.e the snowline is at 3.2 au)\label{fig:snow_ex_2}}
\end{figure*}

Figure \ref{fig:snow_ex_2} shows diameter distributions of  water-ice-rich, organic, and fayalite rich planetesimals at the end of the gas disk phase. The initial size of planetesimals is set  D=100~km and we have performed simulations following the GT, insituJSrapid, and migJSrapid migration histories. Our results for planetesimals of organic and fayalite rich compositions are very similar to those of our nominal case where the disk water snowline is at 1 au rather than 3.3 au. On the other hand, not quite surprisingly,  water-ice-rich planetesimals tend to be more easily destroyed in a hotter disk. As a result, they are more efficiently implanted in the middle and outer regions of the Asteroid Belt (with $a > 2.5$ au, as shown in Figs. \ref{fig:snow_ex_2} (a1), (a2), and (a3)).

\begin{figure*}
    \includegraphics[width=\columnwidth]{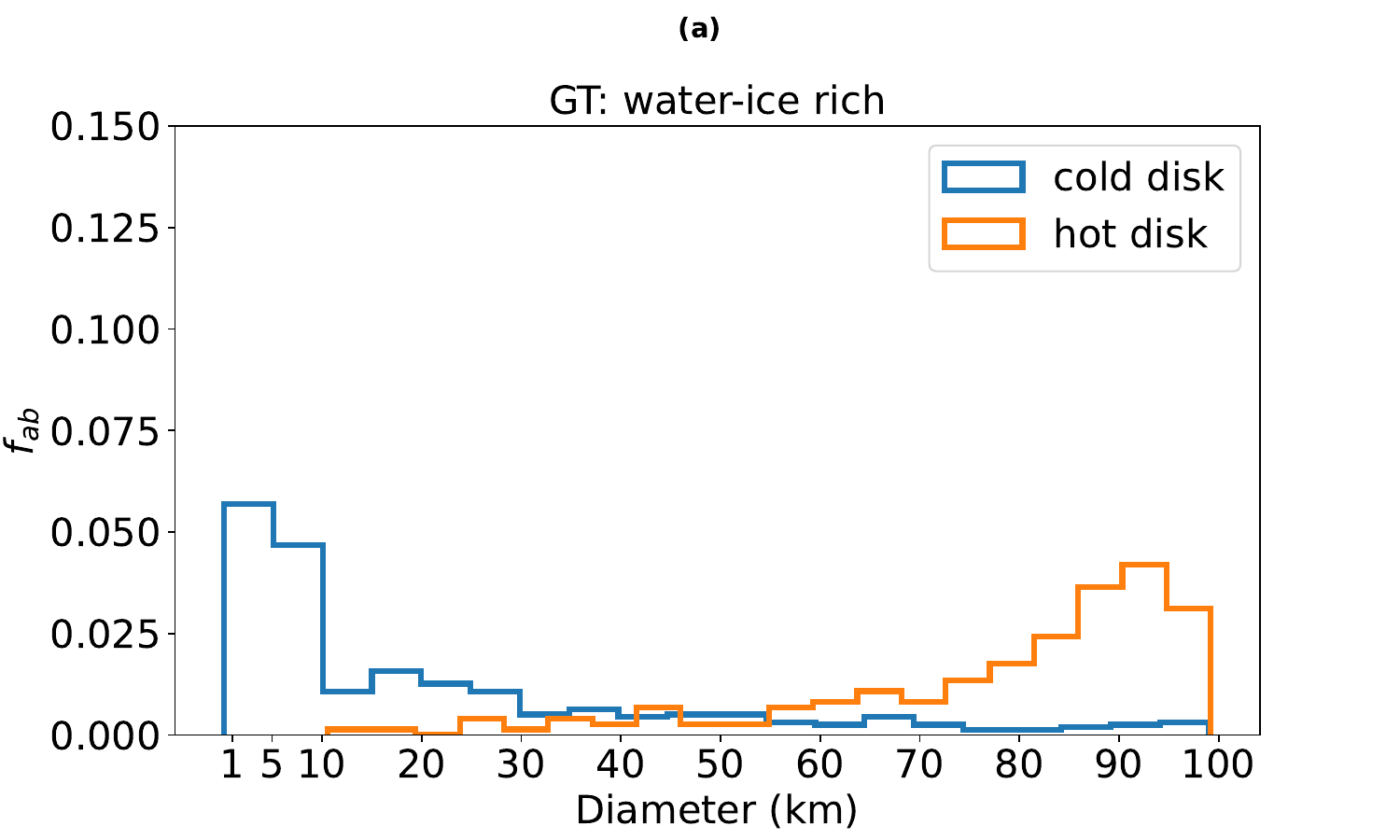}
    \includegraphics[width=\columnwidth]{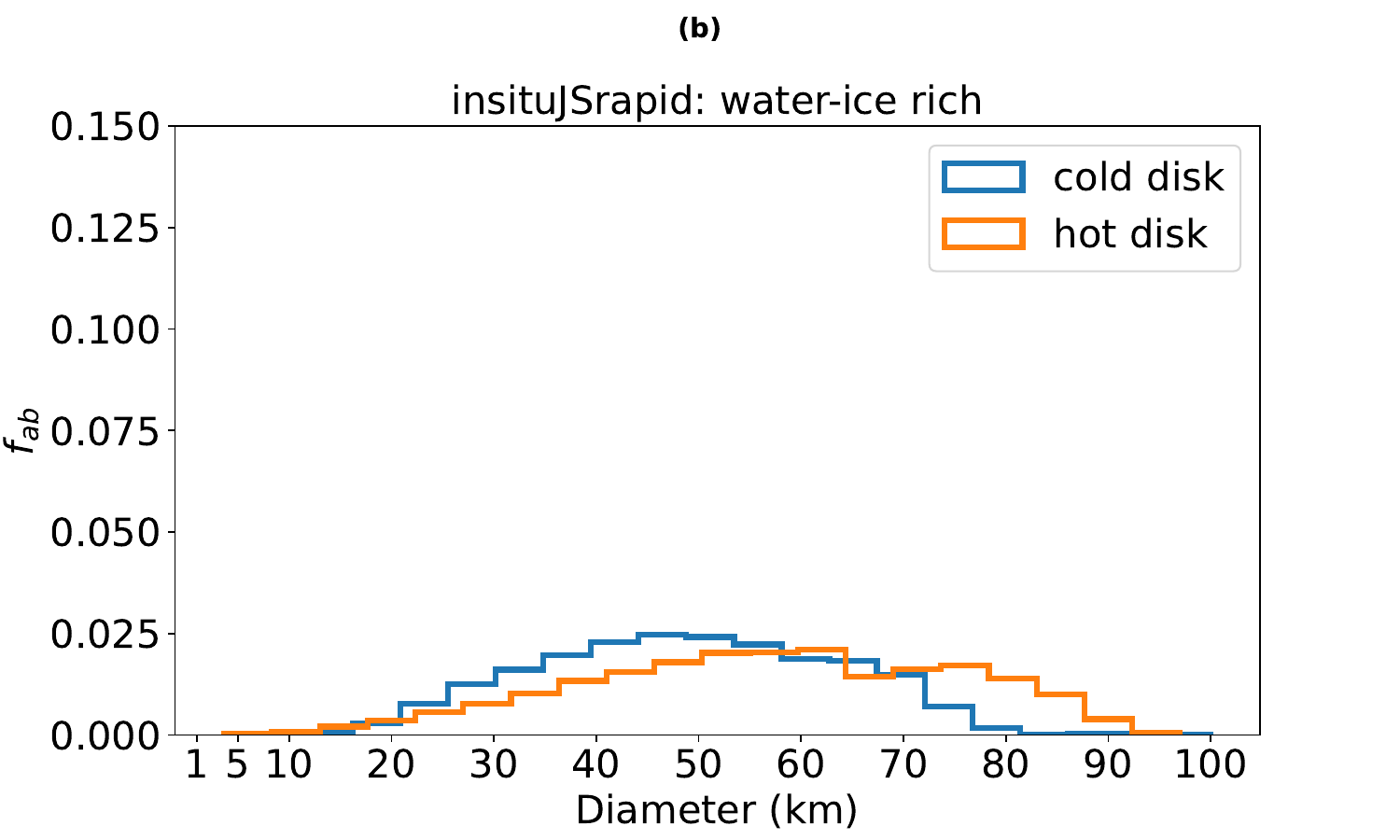}
    \caption{Panels (a) and (b) show histograms depicting the sizes of the initial 100-km water-ice rich planetesimals implanted in the asteroid belt by the GT and insituJSrapid, respectively, at the end of the gas disk phase. The colored lines represent different temperatures of the gas disk: a cold disk (blue line) and hot disks (orange lines). $f_{ab}$ is the relative number of implanted planetesimals.}
    \label{fig:cold_hot}
\end{figure*}

Figure \ref{fig:cold_hot} shows histograms of the size distributions of implanted  water-ice-rich planetesimal in simulations following the GT and insituJSrapid giant planet evolutions. We now compare the results of our cold and hot disk scenarios.  In the cold disk scenario, the GT evolution preferentially implants small planetesimals of D$\approx$5~km. In the hot disk scenario, however, implanted planetesimals have D peaking at about 90~Km instead. This is because planetesimals converted into small planetesimals in the cold disk case are now completely destroyed due to the effects of surface ablation in the hot disk case.
In the insituJSrapid migration scenario, the size distribution of implanted asteroids also shifts towards bigger sizes in the hot disk case, but the impact of the disk temperature is not as strong as in the GT case.

The fraction of water-ice-rich planetesimals completely destroyed by surface ablation in our simulation following the insituJSrapid varies from 8\% to 12\% for cold and hot disks, respectively. In this same simulation, the destruction by ablation accounts for 5\% of organic and fayalite planetesimals for both cold and hot disks. For the GT simulations, regardless of composition, approximately 26\% to 30\% of the planetesimals are completely destroyed by surface ablation for a cold and hot disk, respectively. The inward migration of Jupiter and Saturn during the Grand Tack is the cause of the high efficiency of planetesimal destruction by ablation. The fraction of the overall mass that is lost by ablation, collision, and ejection from the Solar System during the insituJSrapid for water-ice rich planetesimals is about 46\%. This fraction is much bigger for the GT simulations, which is about 81\%.

\subsubsection{The effect of the initial gas surface density}
\label{snowline}

In this section, we show how our results change depending on the initial gas density of the disk.  We have performed simulations considering the insituJSrapid giant planet scenario and planetesimals of water-ice rich compositions. In order to consider different initial gas surface densities, we rescale our nominal and initial gas disk profile by a factor of 0.6, 0.36 or 0.01. We assume that planetesimals have initial sizes of D=100~km.

\begin{table}
	\centering
	\caption{Relative gas surface densities and corresponding fractions of implanted planetesimals into the asteroid belt. We have performed simulations starting with gas disk surface density profiles corresponding to 60\%, 36\% and 1\% of that of our nominal disk model.}
	\label{tab:age_case_1}
	\begin{tabular}{cccc}
		\hline

		  $\Sigma_{o}(r,t)/\Sigma(r,t)$ & fraction of implanted planetesimals \\
		  \hline
		 60\% & $1.1\times 10^{-1}$& \\
		 36\% &  $1.0\times 10^{-1}$  & \\
		 1\% &   $3.0\times 10^{-2}$  & \\
		\hline
	\end{tabular}

	\centering

\end{table}

\begin{figure}
  \includegraphics[width=\columnwidth]{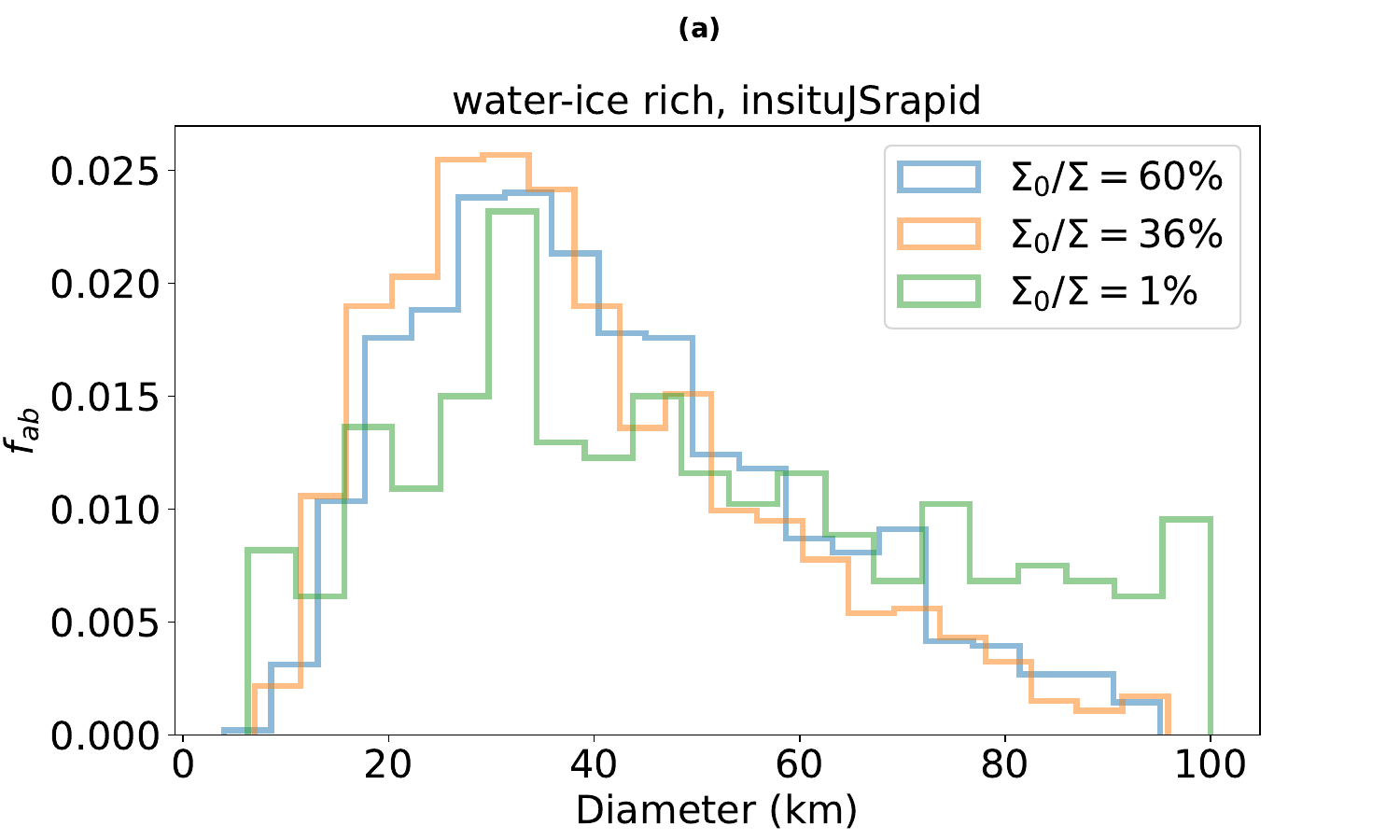}
          \caption{Panels (a) shows histograms of the sizes of the initial 100-km water-ice rich planetesimals implanted in the asteroid belt by the insituJSrapid at the end of the gas disk phase using the relative number of planetesimals. The colored lines represent different initial densities of the gas. \label{fig:density_ablated_gas}}
\end{figure}

 Figure \ref{fig:density_ablated_gas} shows the final size distributions of  planetesimals implanted into the asteroid belt for different initial surface density profiles. Figure \ref{fig:density_ablated_gas}-(a) shows the relative number of planetesimals, whereas Figure \ref{fig:density_ablated_gas}-(b) shows the absolute number of implanted planetesimals.  As one can see, even when our initial disk is significantly gas depleted to 60\% and 36\% of the nominal gas density, surface ablation can significantly reduce the size of water-rich planetesimials. This is also true for our simulations where the initial gas density is reduced to 1\% of that of our nominal disk. However, we caution the reader that this scenario is not realistic because such a low mass disk would not be able to provide enough gas to make the giant planets. Our goal here is to demonstrate that in our model surface mass ablation may also take place in very low density disks sufficiently hot. We note that we keep the disk temperature constant during all our simulations. One of the interesting differences between high and low density disks is that, as shown by Figure \ref{fig:density_ablated_gas}-(b), an initial very low gas density tend to implant fewer planetesimals into the asteroid belt. This result is also consistent with those of  \citet{raymond17}. The fraction of implanted planetesimals into the asteroid belt is given in Table \ref{tab:age_case_1}. Notice also that we are using an exponential decay model for the disk over 100 Ky, which represents a very short timescale. A more slowly decaying disk would induce even more ablation during the implantation process of the planetesimals.

\subsection{Assuming an initial size-frequency distribution for the planetesimal population}
\label{sec:sfd}

Our nominal simulations were conducted assuming populations of equal size planetesimals. In this section, we want to understand how surface ablation may change the initial SFD of a planetesimal population implanted into the asteroid belt. We have performed simulations assuming an initial SFD considering two distinct compositions: water-ice-rich and organic-rich planetesimals. For each planetesimal composition, we create a population of 1.831 million planetesimals  with sizes following a cumulative distribution given by $N(>D) \propto D^{-q}$, where $q=4$ (see Section \ref{sec:initialsize}). Simulations in this section invokes the cold disk scenario.

\begin{figure*}
 \includegraphics[scale=0.27]{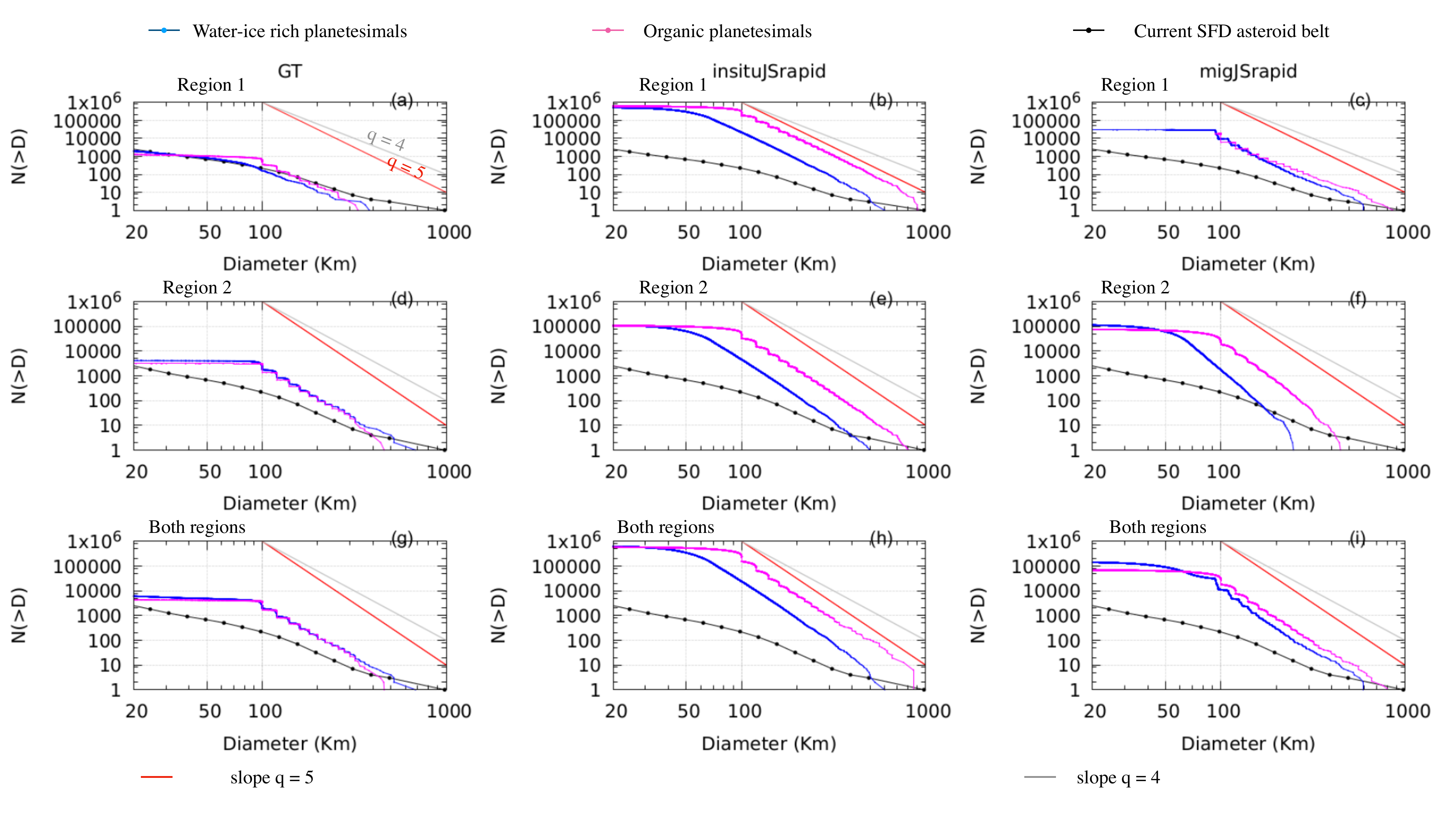}

          \caption{The cumulative size-frequency distribution (SFD) for water-ice-rich (blue line) and organic (magenta line) planetesimals implanted in the asteroid belt. Panels (a), (b), and (c) display the SFD of implanted planetesimals that were initially in Region 1 for each simulation set: GT, insituJSrapid, and migJSrapid, respectively. Center panels (d), (e) and (f) show the SFD of implanted planetesimals for those initially in Region 2 and also for each simulation set. Bottom panels (g), (h), and (i) are for the SFD representing both Regions 1 and 2 combined. Our initial planetesimal population followed a cumulative power slope $q =$ 4, here represented by the gray line. The red line is a reference for the cumulative power slope of $q=$ 5. The current SFD of the asteroid belt is also shown for reference (black). \label{fig:SFDall}}
\end{figure*}

Figure \ref{fig:SFDall} shows the cumulative SFD of water-ice-rich (blue line) and organic (magenta line) planetesimals implanted into the asteroid belt.  Panels (a)-(d)-(g), (b)-(e)-(h), and (c)-(f)-(i) show the final distributions in simulations following the GT, insituJSrapid and migJSrapid  migration histories, respectively. The top and center panels show the final distribution of implanted planetesimals originated from Region 1 and Region 2, respectively. The bottom panel shows the final distribution of implanted planetesimals from both initial regions (Region 1 and Region 2).
In each panel, the solid gray line represents the power slope of the initial planetesimal population. The black solid-dotted line shows SFD of the real asteroid belt for comparison purposes only.

Figure \ref{fig:SFDall} shows that regardless of the dynamical evolution of the giant planets,  planetesimals smaller than 100 km are ``created'' during gas-assisted implantation due to the effects of surface ablation. In our simulations, we do not model the long-term collisional evolution of the asteroid belt \citep{2005Icar..179...63B}, so our goal here is not to match the current asteroid belt SFD but only to understand how surface ablation affects the initial SFD of large implanted asteroids. We are particularly interested in understanding the SFD slope of D$>$100~km planetesimals implanted into the asteroid belt, because these objects are broadly inconsequential to collisional evolution \citep{2005Icar..179...63B}.

Figure \ref{fig:SFDall}-(a) shows that the SFD-slope of planetesimals from Region 1 implanted into the asteroid belt is slightly different  -- shallower -- than the initial one in our GT simulations. However, this is not the case for Region 2. Combined, asteroids implanted from Region 1 and 2 produce a SFD slope (D$>$100~Km) that is qualitatively very similar to that of the initial population (gray line). This is mainly because a relatively larger number of asteroids from Region 2 tend to be implanted into the belt compared to those of Region 1. Interestingly, our simulations following the insituJSrapid and migJSrapid migration scenarios (Figure \ref{fig:SFDall}, Panels (b), (c), (e), (f), (h) and (i)), show that the slope of the implanted planetesimals tend to get slightly steeper relative to that of the initial population (see the red line where q=5). A steeper SFD is mainly due to the fact that large planetesimals are damped less efficiently by gas drag and therefore the velocity difference with respect to the gas is larger than for small bodies, leading to more ablation and increase of the small body population. We note  also that no Ceres-size planetesimals are implanted. This lack of implantation is not attributable to ablation; rather, it is a result of the small implantation probability and the reduced number of initial Ceres-size bodies in our sample of initial conditions.

\begin{figure*}
 \includegraphics[scale=0.27]{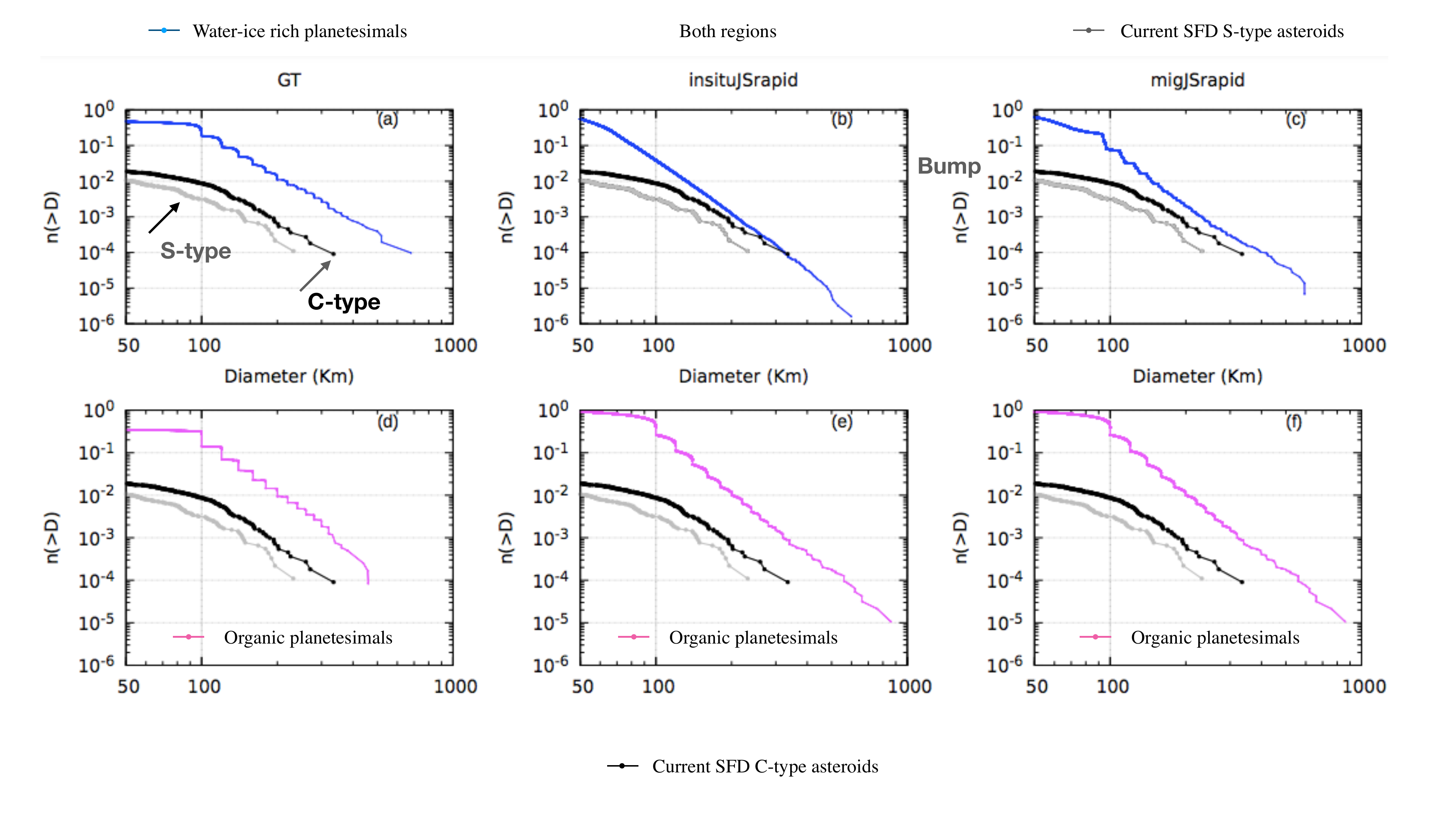}

          \caption{The normalized cumulative size-frequency distribution (SFD) for water-ice-rich (blue line) and organic (magenta line) planetesimals implanted in the asteroid belt. Panels display the SFD of implanted planetesimals that were initially in both regions for each simulation set: GT, insituJSrapid, and migJSrapid, respectively. Panels (a), (b), and (c) show a comparison with the current normalized SFD of the S-type asteroids. Bottom panels (d), (e), and (f) show a comparison with the current normalized SFD of the S and C-types asteroids.\label{fig:NSFD_all}}
\end{figure*}

In Figure \ref{fig:NSFD_all}, we compare the observed size distribution of S-type and C-type asteroids using the Dataset from \citet{DVN/MNA7QG_2017}. From that, we notice that there is no significant difference in the slope between the distributions of S- and C-type asteroids. This is a surprising coincidence given the expected differences in ablation relevance deriving from the different dynamical histories  and compositions of the two populations. However, we cannot be conclusive because of potential observational and taxonomic classification bias \citep{2003Icar..166..116D}.  As we said before, the size distribution of asteroids bigger than 100km is unlikely to have been affected by the collisional evolution that occurred after implantation in the asteroid belt. The currently observed bump in the size distribution at 100km is also unlikely to have been affected \citep{2005Icar..179...63B}. In Figure \ref{fig:NSFD_all}, we notice that in the case where Jupiter and Saturn grow in situ, an initial bump at 100km in the water-rich planetesimal population would have been erased. That could imply such dynamical evolution is not appropriate to explain the SFD of C-type asteroids. On the other hand, it seems more likely that the assumption that the loss of water entrains the removal of any chemical species is not realistic. In the SFD of the organic-rich planetesimals, for example, we see that the location of the bump would not be shifted significantly when considering the same case with Jupiter and Saturn growing in situ.

\section{Conclusions}
\label{conclusion}

In this work, we modeled the implantation of planetesimals into the asteroid belt during the growth and migration of Jupiter and Saturn in their natal disk. Unlike previous studies~\citep{Walshetal2011,raymond17}, our simulations included the effects of surface-mass ablation due to the thermal and frictional heating of planetesimals traveling through the sun's natal gaseous disk.  We have performed a suite of numerical simulations considering different giant planet growth and migration histories. This includes simulations mimicking the evolution of Jupiter and Saturn in the so-called Grand-Tack model~\citep{Walsh2011}, and also simulations where these planets form nearly in-situ, and a scenario  where they are assumed to experience only inward migration~\citep{raymond17}. Our simulations  account for different disk temperature profiles, disk density profiles, planetesimals sizes, and planetesimal compositions. We studied the effects of surface ablation on  water-ice rich, water-ice poor (enstatite silicates), organic-rich, and fayalite-rich planetesimals.

Our results suggest that regardless of the migration histories of the giant planets, gas disk radial density profile, gas disk temperature radial profile, planetesimal composition and size, planetesimals implanted into the asteroid belt during the growth and migration of Jupiter and Saturn are very likely to have experienced surface ablation and decreased in size during the implantation process. Only planetesimals with water-ice poor compositions  -- made of enstatite minerals --  were largely resistant to heating and remained  unaffected by surface ablation being, consequently, implanted into the asteroid belt with their original sizes.

We also analyzed how surface ablation during gas-assisted implantation changes the initial size-frequency distribution of planetesimals implanted into the asteroid belt. We found that surface mass ablation tend to make the SFD slope slightly steeper than the original population, by reducing the number of relatively larger planetesimals and increasing the number of relatively smaller ones during the implantation process.  Therefore, our results  (cold disk scenario) suggest that the initial SFD slope of the planetesimal population implanted into the asteroid belt is slightly different from that of the initial planetesimal population.  With the exception of the case where planetesimals were enstatite-like (making the effects of surface ablation are negligible), the above conclusion holds true for all other implantation scenarios considered in our simulations.

Recent solar system formation models suggest that asteroids from the inner (S-type) and outer (C-type) solar system were implanted into the asteroid belt via different dynamical mechanisms~\citep{raymond17,2017SciARaymondIzidoro}. While the implantation of C-type asteroids is assisted by gas-drag, as modelled here, the implantation of S-type asteroids is suggested to be an outcome of gravitational scattering in a gas-free scenario~\citep{2017SciARaymondIzidoro,2021NatAs.tmp..262I}. If this view is correct, it implies that surface ablation did not affect S-type asteroids during the implantation process.

Finally, our model is admittedly simplified in many aspects. For instance, our planetesimals are considered to be made of a pure chemical compound. In addition, our surface ablation model neglects many important processes as the effects of fragmentation and collisional evolution. Yet, our study demonstrate, for a reasonable range of parameter space, that surface ablation of planetesimals may be an important process in protoplanetary disks, in particular for water-ice rich planetesimals. We acknowledge that the efficiency of surface mass ablation strongly depends  on the planetesimal composition and surface material ``strenght'', parameters yet poorly constrained from both analytical and observational grounds. Our results, however, suggest that surface mass ablation deserves further attention and investigation in dynamical models of planet formation. For instance, our simulations suggest that high-level surface ablation may have even exposed the cores of early differentiated planetesimals implanted into the asteroid belt, and that asteroids were likely born bigger than currently envisioned. Future observations and asteroid missions should look for signposts of the effects of ablation in the minor bodies of the asteroid belt.

\section*{Acknowledgements \label{Sec:Acknowledgements}}
We thank Alessandro Morbidelli and the anonymous referee for the valuable comments and suggestions that helped improve the quality of this paper. Rafael Ribeiro (RR) thanks to the scholarship granted from the Brazilian Federal Agency for Support and Evaluation of Graduate Education (CAPES), in the scope of the Program CAPES-PrInt (Proc~88887.310463/2018-00, Mobility number 88887.572647/2020-00, 88887.468205/2019-00). This research was supported in part by the São Paulo Research Foundation (FAPESP) through the computational resources provided by the Center for Scientific Computing (NCC/GridUNESP) of the São Paulo State University (UNESP). RR also  acknowledges the support provided by grants FAPESP (Proc~2016/24561-0) and by Sao Paulo State University (PROPe~13/2022).  R.D. was supported by the NASA Emerging Worlds program, grant 80NSSC21K0387. A.I. and R. Dasgupta acknowledge NASA grant 80NSSC18K0828 during preparation and submission of the work.



%
\bibliographystyle{aasjournal}

\bibliography{example} 

\end{document}